\documentclass[aps,prl,onecolumn,showpacs,preprintnumbers]{revtex4}
\usepackage{amsmath}
\usepackage{dcolumn}
\usepackage{bm}
\usepackage{subfigure}
\usepackage{amsfonts}
\usepackage{amssymb}
\usepackage{makeidx}
\usepackage{epsfig}
\usepackage{graphicx}

\begin{document}

\title[]{Theory of non-local point transformations - Part 1: \
Representation of Teleparallel Gravity}
\author{Massimo Tessarotto}
\affiliation{Department of Mathematics and Geosciences, University of Trieste, Italy}
\affiliation{Institute of Physics, Faculty of Philosophy and Science, Silesian University
in Opava, Bezru\v{c}ovo n\'{a}m.13, CZ-74601 Opava, Czech Republic}
\author{Claudio Cremaschini}
\affiliation{Institute of Physics, Faculty of Philosophy and Science, Silesian University
in Opava, Bezru\v{c}ovo n\'{a}m.13, CZ-74601 Opava, Czech Republic}
\date{\today }

\begin{abstract}
In this paper the extension of\ the functional setting customarily adopted
in General Relativity (GR) is considered. For this purpose, an explicit
solution of the so-called Einstein's\ Teleparallel problem is sought. This
is achieved by a suitable extension of the traditional concept of GR
reference frame and is based on the notion of non-local point transformation
(NLPT). In particular, it is shown that a solution to the said problem can
be reached by introducing a suitable subset of transformations denoted here
as \textit{special} \textit{NLPT}. These are found to realize a phase-space
transformation connecting\emph{\ }the flat Minkowski space-time with, in
principle, an arbitrary curved space-time. The functional setting and basic
properties of the new transformations are investigated.
\end{abstract}

\maketitle

%\affiliation{}

%\tableofcontents

%\frontmatter

%\email{cremasch@sissa.it}

%\email{M.Tessarotto@cmfd.univ.trieste.it}

\section{1 - Introduction}

In this paper and in the subsequent related ones (Parts 2 and 3) the problem
is investigated of\ the extension of\ the customary functional setting which
lays at the\ basis of relativistic\ theories in physics. These notably
include, besides classical electrodynamics, relativistic classical mechanics
and relativistic quantum mechanics, in particular the so-called \emph{%
Standard Formulation to General Relativity} (SF-GR), i.e., Einstein's
original approach to his namesake field equations. The latter, as is well
known, uniquely determine the metric tensor $g_{\mu \nu }\left( r\right) $
\cite{Einstein1915,Landau,Wheeler,Synge,wald,noi2015} associated with a
prescribed parametrization of the physical space-time, identified with the $%
4-$dimensional connected and time-oriented real metric space $D^{4}\equiv
\left( \mathbf{Q}^{4},g\right) $, with $\mathbf{Q}^{4}\equiv
%TCIMACRO{\U{211d} }%
%BeginExpansion
\mathbb{R}
%EndExpansion
^{4}$.\textbf{\ }Such a functional setting is realized by the group of
transformations connecting arbitrary GR-reference frames, i.e., arbitrary $%
4- $dimensional curvilinear coordinate systems spanning the same prescribed
space-time $D^{4}$.\textbf{\ }In SF-GR this is usually identified with the
group\textbf{\ }($\left\{ P\right\} $) of invertible\textbf{\ }\emph{local
point transformations (LPT)} $P$ and its inverse $P^{-1}$, namely%
\begin{eqnarray}
P &:&r^{\mu }\rightarrow r^{\prime \mu }=r^{\prime \mu }(r),  \label{T-1} \\
P^{-1} &:&r^{\prime \mu }\rightarrow r^{\mu }=r^{\mu }(r^{\prime }),
\label{T-2}
\end{eqnarray}%
where the initial and transformed $4-$positions $r^{\mu }$\ and\ $r^{\prime
\mu }$ are assumed to span the same space-time $(\mathbf{Q}^{4},g)$. Hence,
by definition, the group $\left\{ P\right\} $ leaves invariant $(\mathbf{Q}%
^{4},g)$, which must therefore be identified with a differential manifold.
It is obvious that such a functional setting is intrinsic to SF-GR, i.e., it
is actually required for the validity of SF-GR itself. The same
transformations\textbf{\emph{\ }}(Eqs.(\ref{T-1})-(\ref{T-2})) are assumed
also to warrant the global validity of the so-called Einstein's \emph{%
General Covariance Principle} (GCP) \cite{Einst}. In other words, the
transformations defined by Eqs.(\ref{T-1})-(\ref{T-2}) must be endowed with
a suitable\ functional setting (see related discussion in Section 2),
referred to here as\textbf{\emph{\ }}\emph{LPT-functional setting}, which
permits in turn also the\ corresponding realization of GCP. Such a principle
is therefore referred to as\ \emph{LPT-GCP}.\textbf{\ }In particular, this
means that LPT must be smoothly differentiable so to uniquely and globally
prescribe also the $4-$\emph{tensor transformation laws} of the displacement
$4-$vectors, namely%
\begin{equation}
\left\{
\begin{tabular}{l}
$dr^{\mu }=\mathcal{J}_{\nu }^{\mu }dr^{\prime \nu },$ \\
$dr^{\prime \mu }=\left( \mathcal{J}^{-1}\right) _{\nu }^{\mu }dr^{\nu }.$%
\end{tabular}%
\right.  \label{dT-1}
\end{equation}%
Here $\mathcal{J}_{\nu }^{\mu }$\ and $\left( \mathcal{J}^{-1}\right) _{\nu
}^{\mu }$ denote the\ direct and inverse Jacobian matrices which take the
so-called \emph{gradient form}, i.e.,
\begin{eqnarray}
\mathcal{J}_{\nu }^{\mu }(r^{\prime }) &\equiv &\frac{\partial r^{\mu
}(r^{\prime })}{\partial r^{\prime \nu }},  \label{GR-1} \\
\left( \mathcal{J}^{-1}\right) _{\nu }^{\mu }(r) &\equiv &\frac{\partial
r^{\prime \mu }(r)}{\partial r^{\nu }},  \label{GR-2}
\end{eqnarray}%
which uniquely-globally prescribe also the corresponding $4-$tensor
transformation laws\emph{\ }of \emph{all tensor fields} which characterize
SF-GR.

However, in this work we intend to show that a new approach alternative to
the one adopted in GR\textbf{\ }founded on the introduction of an extended
functional setting\textbf{\ }is actually possible. This is based both on
mathematical and physical considerations.\ Starting point is the notion of
\emph{non-local point transformations (NLPT)}, and will\ be referred to here
as\textbf{\ }\emph{NLPT-functional setting.}\textbf{\ }Such a setting should
permit, in principle, to map in each other intrinsically different
space-times\textbf{\ }$\left( \mathbf{Q}^{4},g\right) $ and $\left( \mathbf{Q%
}^{\prime 4},g^{\prime }\right) $, i.e., space-times which cannot be
otherwise connected by means of the group $\left\{ P\right\} $.

The issue concerns the prescription of the appropriate class of GR-reference
frames (\emph{GR-frames}) to be adopted as well as of the transformations
connecting them. It is well-known that in the customary approach to GR%
\textbf{\ }\cite{Einstein1915,Landau,Wheeler,Synge,wald} the GR-frames are
identified with arbitrary sets of curvilinear coordinate systems, while the
latter are realized by means of LPT, i.e., smoothly differentiable real maps
depending locally on position only. Nevertheless, as discussed below, there
exist theoretical motivations which suggest the mathematical and/or physical
inadequacy (in the context of GR) equivalently either of the functional
setting based on LPT only or\ the traditional concept of GR reference frame,%
\textbf{\ }which in fact relies - in turn - on the use of the same type of
coordinate transformations. These motivations include a number of
problem-cases of special physical relevance (see below).

In Part 1, in particular, the example-case is considered which deals with
the so-called teleparallel representation of GR, also known as \emph{%
Einstein Teleparallelism}\ or (Einstein)\emph{\ Teleparallel Gravity} \cite%
{ein28}. We intend to prove that in the context of teleparallel gravity the
introduction of new types of GR-frames and coordinate transformations is
mandatory.\ These are found to be realized respectively by means of a kind
of phase-space reference frames, denoted as \emph{extended GR-frames}, and a
suitably-defined set of phase-space maps, which involve in particular the
introduction of appropriate non-local coordinate transformations, identified
here as \emph{special NLPT.}

\subsection{Historical ante factum and the issue of non-local
generalizations of GR}

An ongoing subject of theoretical investigations in GR concerns its possible
non-local modifications. Recent literature investigations in this category
are several. Examples can be found, for instance, in Refs.\cite%
{h0,h1,h2,h3,h6,h4}, where non-local generalizations of the Einstein theory
of gravitation have been proposed.\textbf{\ }Such a kind of non-local GR
models lead typically to suitably-modified forms of the Einstein equation
\cite{Einstein1915} in which non-local field interactions are accounted for,
in analogy with corresponding non-local features of the electromagnetic
field occurring in Classical Electrodynamics.

It is well-know that the \emph{LPT-functional setting} adopted by
Einstein in his original formulation of GR is uniquely
founded on the \textbf{\ }classical theory of tensor calculus on manifolds%
\textbf{.} The historical foundations of the latter, in turn, date
back to the so-called absolute differential calculus developed at
the end of 19th-century by Gregorio Ricci-Curbastro and later
popularized by his former student and collaborator Tullio
Levi-Civita \cite{Wheeler2,Synge}. \ However, a basic issue that
arises in GR and its possible non-local generalizations, as well
as more generally in classical and quantum theories of particles
and fields, is whether these theories themselves might exhibit
possible contradictions with the validity of the LPT-GCP and
consequently a more general functional setting should be actually
adopted for the treatment of these disciplines.

To better elucidate the scope and potential physical relevance of the topics
indicated above, it is worth to highlight in detail some of the main related
issues and physical problems to be found in the literature which, as
explained in detail below, are still challenging and whose solution appears
of critical importance in GR. These include:

\begin{enumerate}
\item \emph{Problem \#1: Teleparallel approach to GR - }One of the most
remarkable physical examples of violation of LPT--GCP - and the one which
motivates the present paper - occurs however in the framework of the
Einstein's teleparallelism\ (or Teleparallel problem, see Refs.\cite{ein28}%
), and possibly also in some of its recently-proposed generalizations \cite%
{tele1,tele2,tele3}.\textbf{\ }The conclusion is of immediate and patent
evidence. Indeed, such a theory is intended to map in each other
intrinsically different space-times. In the case of Teleparallelism one of
such space-times is identified, by construction, with the flat time-oriented
Minkowski space-time.\ As discussed below (see Section 3), this is achieved
by a suitable matrix transformation (\emph{teleparallel transformation})
between the corresponding metric tensors, denoted as teleparallel problem
(TT-problem), which lies at the basis of such an approach (see Eq.(\ref%
{first equation}) or equivalent Eq.(\ref{last-equation})). A number of
related issues arise which concern in particular:

\begin{itemize}
\item \emph{Problem \#P1}$_{1}$\emph{\ }- The realization and possible
non-uniqueness feature of the mapping to be established between the two
space-times occurring in the teleparallel transformation itself. This refers
in particular of what might/should be:

A) the actual representation of the corresponding coordinate transformations;

B) their local and possible non-local dependences;

C) the possible existence/non-existence of corresponding tensor
transformation laws for observable tensor fields, etc.

\item \emph{Problem \#P1}$_{2}$ - The fact that obviously such problems, and
the TT-problem itself, cannot be solved in the framework of the validity of
the LPT-GCP.

\item \emph{Problem \#P1}$_{3}$ - The physical implications of the theory,
with particular reference to\ the explicit construction of special NLPT.
\end{itemize}

\item \emph{Problem \#2: Diagonalization of metric tensors and complex
transformation approaches to GR - }A second notable example concerns the
adoption in GR of complex-variable transformations, such as the so-called
Newman-Janis algorithm \cite{NJ1,NJ2,NJ3}. This is frequently used in the
literature for the purpose of investigating a variety of standard or
non-standard GR black-hole solutions \cite{bambi,bambi2}, as well as
alternative theories of gravitation, such as the one based on
non-commutative geometry \cite{Nicolini-nj}.\textbf{\ }Its basic feature is
that of permitting one to transform, by means of a complex coordinate
transformation, a diagonal metric tensor corresponding to a
spherically-symmetric and stationary configuration (like the Schwarzschild
one) into a non-diagonal one corresponding to a rotating black-hole (like
the Kerr solution). On the other hand, a number of issues arise concerning
the Newman-Janis algorithm. These include:

\begin{itemize}
\item \emph{Problem \#P2}$_{1}$ - First, it is complex, so that the
transformed coordinates are complex too. This inhibits their objective
physical interpretation in terms of physical observables.

\item \emph{Problem \#P2}$_{2}$ - The fact that, as for the Teleparallel
transformation, the diagonalization problem at the basis of the same
transformation cannot be solved in the framework of the validity of the
LPT-GCP. Indeed, the Newman-Janis algorithm seems worth to be mentioned
especially in view of the fact that it obviously represents a patent
violation of the LPT-GCP.

\item \emph{Problem \#P2}$_{3}$ - The physical meaning of the
transformation: one cannot ignore that fact that there is no clear
understanding regarding its physical interpretation and ultimately as to why
the algorithm should actually work at all.

\item \emph{Problem \#P2}$_{4}$ - Finally, despite the obvious fact that the
Teleparallel transformation provides in principle also a solution to the
diagonalization problem, there is no clear connection emerging between the
same transformation and the Newman-Janis algorithm.
\end{itemize}

\item \emph{Problem \#3: Acceleration effects in relativistic classical
electrodynamics -\ }A third issue worth to be pointed out for its potential
relevance in the present discussion concerns the role of acceleration on GR
reference frames as discussed for example in Refs.\cite{mash1,mash2}. These
papers deal with the necessity of taking into account, both in the context
of GR and Maxwell's equations, possible acceleration-induced non-local
effects. However, the precise mathematical formulation and physical
mechanisms by which non-locality should manifest itself must still be fully
understood.

In fact, a number of basic issues remain unanswered. These concern in
particular the following ones:

\begin{itemize}
\item \emph{Problem \#P3}$_{1}$ - First, the precise prescription of the
mathematical setting of the theory and in particular the implementation and
possible\ functional realization of the non-local acceleration effects and
the possible connection with the theory of Teleparallel gravity in the
context of GR remain unclear.

\item \emph{Problem \#P3}$_{2}$ - Indeed, non-local acceleration effects are
introduced by postulating directly "ad hoc" integral representations (or
"transformation laws") for appropriate tensor fields.

\item \emph{Problem \#P3}$_{3}$ - The validity of these transformation laws,
namely the reason why ultimately they should apply, and consequently their
physical interpretation, remain both ultimately unclear.
\end{itemize}

\item \emph{Problem \#4: Non--local effects in classical electrodynamics - }%
A further intriguing example which is by itself sufficient to demonstrate
the role of non-locality in physics can be found in the framework of a
special-relativistic treatment of classical electrodynamics. This concerns
the so-called electromagnetic radiation-reaction (EM-RR) problem, i.e., the
dynamics of an extended charge in the presence of its self-generated EM
field. As shown in Refs.\cite{EPJ1,EPJ2} such a problem can be rigorously
treated in the framework of a first-principle approach based on the Hamilton
variational principle. In such a context the sources of non-locality appears
at once as being due to the finite size of charged particles. Indeed, its
physical origin is related to the retarded EM interaction of the extended
particle with itself \cite{EPJ3,EPJ5,EPJ6,EPJ7,EPJ8}.\textbf{\ }However, a
further fundamental physical implication also emerges.\textbf{\ }In fact, as
shown in Ref.\cite{EPJ3}, in the variational action functional \emph{point
Lorentz transformations must be considered as non-local,} thus effectively
extending the class of local Lorentz transformations usually considered in
special relativity. This arises because, in order to preserve the scalar
property of the relativistic Lagrangian in the Hamilton variational
principle, the point transformations (realized by Lorentz transformations)
must act \textquotedblleft non-locally\textquotedblright . In fact, in
contrast to local coordinate transformations which are supposed to act only
on the explicit local functional dependences, in such a case the Lorentz
transformations must also act on the non-local dependences appearing in the
same functional. In particular, the following issues should be answered:

\begin{itemize}
\item \emph{Problem \#P4}$_{1}$ - First, the precise prescription of the
transformation laws with respect to the group on NLPT should be achieved for
the EM $4-$potential $A^{\mu }$ and of the corresponding EM Faraday tensor $%
F^{\mu \nu }$.

\item \emph{Problem \#P4}$_{2}$ - Second, it remains to be ascertained
whether and possibly under what conditions the transformations indicated
above are realized by means tensor transformation laws, i.e., respectively
for $A^{\mu }$\ and $F^{\mu \nu }$, transformation laws formally identical
to those determined by the $4-$position infinitesimal displacement $dr^{\mu
} $\ or the dyadic tensor $dr^{\mu }dr^{\nu }.$
\end{itemize}
\end{enumerate}

The key question which needs to be ascertained in the context of GR is
whether these problems do actually require, as anticipated above, the
introduction of a more general class of GR-reference frames. In fact,
despite previous interesting but incomplete solution attempts \cite%
{mash1,mash2}, a basic issue which still remains unsolved nowadays concerns
the construction of the explicit general form and physically-admissible
realizations which the transformations occurring among arbitrary GR-frames
should take. The problem matter refers therefore to possible non-local
generalization of the customary local tensor calculus and coordinate
transformations to be adopted in GR.\textbf{\ }This is actually the task
which we intend to undertake in the present investigation.

Under such premises it must be noted that the present work departs, while
being at the same time also in some sense complementary, from the non-local
GR theories indicated above. In fact it belongs to the class of studies
aimed at introducing in the context of GR a new type of non-local phenomena
based on the coordinate transformations established between GR-reference
frames and at the same time extending the functional setting customarily
adopted in such a context.

\subsection{Outline of the investigation}

More precisely, the overall work-plan of the investigation is to address the
problem of the non-local generalization of GR achieved by a suitable
extension of its functional setting. This task is by no means trivial since
it concerns basic theoretical issues and physical problems which have
remained unsolved to date in the literature and whose solution presented in
this investigation for the first time appears of critical importance in
General Relativity (GR). In detail these include:

\begin{enumerate}
\item \emph{Topics \#1} - The identification of possible generalizations of
the LPT-setting customarily adopted in GR, based on physical example-cases.
A notable problem of this type is realized by Einstein's approach to the
so-called \emph{Einstein's teleparallelism}. In such a context, the issue
arises whether such a theory can be recovered from SF-GR by means of a
suitable mathematical\emph{,} i.e., purely conceptual, viewpoint. This
involves the introduction of appropriate non-local point transformations (or
NLPT). Their determination, despite being of basic importance in GR, still
remains essentially unknown to date. It must be stressed, in this regard,
that the possible prescription of NLPT is by no means "\textit{a priori"}
obvious since they remain - it must be stressed - largely arbitrary and
intrinsically non-unique. For this purpose in Part 1 \emph{Problems} \emph{%
\#P1}$_{1}-$\emph{\#P1}$_{3}$ are addressed. Their solution is crucial for
their identification. This goal can be reached based on the adoption of a
suitable sub-set of NLPT, referred to here as \emph{special NLPT-group }$%
\left\{ P_{S}\right\} $ acting on appropriate \emph{extended GR-frames}
which are defined with respect to prescribed space-times. For definiteness,
in view of warranting the validity of suitable tensor transformation laws
for the metric tensor which is associated with the Teleparallel
transformation (see Eq.(\ref{transf-2}) below), in the present treatment
these transformations are assumed to preserve the line element (see Section
4 below), in other words the are required to map space-times\textbf{\ }$(%
\mathbf{Q}^{4},g)$ and $(\mathbf{Q}^{\prime 4},g^{\prime })\equiv (\mathbf{M}%
^{\prime 4},\eta )$ having the same line elements\textbf{\ }$ds$ and $%
ds^{\prime }$.

\item \emph{Topics \#2} - In Part 2\textbf{\ }\emph{Problems \#P2}$_{1}-\#$%
\emph{P2}$_{4}$ are addressed. For such a purpose the determination is done
of the group of \emph{general non-local point transformations} (general
NLPT) connecting subsets of two generic curved space-times $(\mathbf{Q}%
^{4},g)$ and $(\mathbf{Q}^{\prime 4},g^{\prime }).$\textbf{\ }This is
referred to here as\textbf{\ }\emph{general NLPT-group }$\left\{
P_{g}\right\} .$ The task posed here involves also their physical
interpretation based on a suitable Gedanken experiment. This refers, in
particular to three distinct issues:

A) The possible conceptual realization of a measure experiment (Gedanken
experiment), simulating the action of a generic, NLPT on a GR-reference
frame on the physical space-time.

B) The prescription of the family of NLPT, exclusively based on a suitable
set of mathematical, i.e., axiomatic, prescriptions, which should be
nevertheless physically realizable in principle for arbitrary GR-reference
frames which are defined with respect to a prescribed (physical) space-time.

C) As an illustration of the theory, the explicit construction of possible
physically-relevant transformations of the group $\left\{ P_{g}\right\} ,$
with special reference to the problem of the \emph{diagonalization of metric
tensors }in GR.

\item \emph{Topics \#3} - The investigation of \emph{physical implications}
of the general NLPT-functional setting, with particular\ reference to the
identification of possible \emph{acceleration effects} in GR and classical
electrodynamics. The goal of Part 3 is to look for a possible solution of
\emph{Problems \#3 and \#4} indicated above. This involves in particular:

A) the investigation of the role of acceleration on GR reference frames;

B) the search of possible $4$-tensor transformation laws occurring
respectively for the $4-$acceleration field and the EM $4-$vector potential
will be investigated,\ with respect to the group of NLPT $\left\{
P_{g}\right\} $ established between suitable subsets of two arbitrary curved
space-times\textbf{\ }$(\mathbf{Q}^{4},g)$ and $(\mathbf{Q}^{\prime
4},g^{\prime })$. Regarding point B), the key related issue concerns in fact
to ascertain whether and under what conditions $4-$tensor transformation
laws exist both for the $4-$acceleration and the EM $4-$vector potential.
\end{enumerate}

In the present manuscript (Part 1) topics \#1 will be addressed. Topics \#2
and \#3 will be, instead, discussed respectively in Parts 2 and 3.

\subsection{Goals and structure of the paper}

Given these premises, we are now in position to state in detail the
structure of the present manuscript, pointing out the goals posed in each of
the following sections which are accordingly listed below.

\begin{enumerate}
\item \emph{GOAL \#1 - }The first one, discussed in\textbf{\ }Section 2,
includes the task of displaying the functional setting (LPT-functional
setting) usually adopted in SF-GR. Its basic features are pointed out
together with some basic implications relevant in the subsequent discussion.

\item \emph{GOAL \#2 - }The second one, which is presented in\ Section 3,%
\textbf{\textbf{\ }}concerns an insight of the Einstein's theory of
teleparallelism and the related Teleparallel Problem (\emph{TT-problem}).
For this purpose its basic assumptions, formulation and implications are
analyzed in detail.

\item \emph{GOAL \#3 - }Based on the investigation of the same TT-problem,
in Section 4 the theory of \emph{special NLPT} is developed. It is shown
that for this purpose a new \emph{NLPT-functional setting} is required. As a
consequence it is shown that a phase-space map can be established between
the Minkowski flat space-time and an in principle arbitrary curved
space-time. This involves, in particular, the adoption of non--local point
transformation, referred to as special NLPT.

\item \emph{GOAL \#4 - }In Section 5, the conditions of existence of NLPT
are discussed, which yield particular solutions of the TT-problem.

\item \emph{GOAL \#5 - }In Section 6, as application of the theory of
special NLPT, a sample case is investigated.

\item \emph{GOAL \#6 - }Finally, in Section 7 the main conclusions of the
paper are drawn.
\end{enumerate}

\section{2 - The LPT- functional setting and its implications}

In order to state clearly the problem and its related motivations, we first
recall the functional setting which - as anticipated above - is usually
adopted both in relativistic theories as well as in Einstein's 1915 theory
of gravitation \cite{Einstein1915}, i.e., SF-GR itself. In both cases the
goal is, in principle, to predict all physically-relevant realizations of
the observables. In the case of GR these concern the physical space-time
itself $D^{4}\equiv \left( \mathbf{Q}^{4},g\right) $.\textbf{\ }As is
well-known, in SF-GR this is identified with a $4-$dimensional Lorentzian
metric space on\textbf{\ }$\mathbf{Q}^{4}\equiv \mathbf{%
%TCIMACRO{\U{211d} }%
%BeginExpansion
\mathbb{R}
%EndExpansion
}^{4}$\textbf{\ }which is endowed with a prescribed metric tensor $g_{\mu
\nu }\left( r\right) $\textbf{\ }when the same set $Q^{4}$\ is represented
in terms of a given set of curvilinear coordinates $\left\{ r^{\mu }\right\}
\equiv r$.\textbf{\ }Nevertheless, validity of GR, and in particular of the
Einstein equation itself, requires to couch them in a suitable mathematical
framework.

As recently pointed out in Ref.\cite{noi2015} in the context of a
variational treatment of SF-GR, this involves, besides the fulfillment of a
suitable property of gauge invariance, also the adoption of Classical Tensor
Analysis on Manifolds. In other words, as anticipated above, both GR and the
same Einstein equation should embody by construction the validity of
LPT-GCP, namely formulated consistent\textbf{\ }with the LPT- functional
setting. This means explicitly that the following mathematical requirements (%
\emph{A-C}) should apply:

\emph{A)} All physically-observable tensor fields defined on space-time $%
\left( \mathbf{Q}^{4},g\right) $ must be realized by means of\emph{\ }$4-$%
tensor fields with respect to a suitable ensemble of coordinate
transformations connecting in principle arbitrary, but suitably related, $4-$%
dimensional curvilinear coordinate systems, referred to as \emph{%
GR-reference frames}, $r^{\mu }$\emph{\ }and\emph{\ }$r^{\prime \mu }$.

\emph{B)} The PDEs, together with their corresponding variational
principles, which characterize all classical and quantum physical laws
should satisfy\emph{\ the criterion of manifest covariance, whereby it
should be possible to cast them in all their realizations in manifest }$4-$%
\emph{tensor form.}

\emph{C)} The set of coordinate transformations indicated above is
identified with the group of transformations that in Eulerian form are
prescribed by means of the invertible maps (\ref{T-1})-(\ref{T-2}) which
identify the group $\left\{ P\right\} $.\emph{\ }For this purpose, suitable
restrictions must be placed on the admissible GR-reference frames, i.e.,
coordinate systems, prescribed by means of Eqs.(\ref{T-1}) and (\ref{T-2})
which are realized by\ the following requirements:

\begin{itemize}
\item \emph{LPT-requirement \#1 - }For the validity of GCP, the two
space-times must coincide and be transformed in one another by means of LPT,
i.e., $(\mathbf{Q}^{4},g(r))$\ $\equiv $ $(\mathbf{Q}^{\prime 4},g^{\prime
}(r^{\prime }))$, so that to define a single $C^{k}-$\emph{differentiable
Lorentzian manifold} with $k\geq 3,$ i.e., have either signature $(+,-,-,-)$
or analogous permutations.

\item \emph{LPT-requirement \# 2 -} These transformations must be assumed as
purely local, so that in Eqs.(\ref{T-1}) and (\ref{T-2}) $r^{\prime \mu }$
and $r^{\mu }$ must depend only locally respectively on $r\equiv \left\{
r^{\mu }\right\} $ and $r^{\prime }\equiv \left\{ r^{\prime \mu }\right\} $.
In other words, the \emph{local values} $r^{\mu }$ and $r^{\prime \mu }$ are
required to be mutually mapped in each other by means of the same equations,
with $r^{\prime \mu }$ (respectively $r^{\mu }$) being a function of $r^{\mu
}$ (and similarly\ $r^{\prime \mu }$) \emph{only}.

\item \emph{LPT-requirement \#3 - }The coordinates $r^{\mu }$\ and\ $%
r^{\prime \mu }$\ must realize \emph{physical observables} and hence be
prescribed in terms of real variables, while the functions relating them ($P$
and $P^{-1})$ must be suitably smooth in the sense that they are of class $%
C^{(k)}$,\ with $k\geq 3$. This means that $(\mathbf{Q}^{4},g(r))$ must
realize a $C^{k}-$\emph{differentiable Lorentzian manifold} with $k\geq 3$.

\item \emph{LPT-requirement \#4 - }Eqs.(\ref{T-1}) and \ref{T-2}) generate
the corresponding $4-$\emph{vector transformation equations} for the
contravariant components of the displacement $4-$vectors $dr^{\mu }$ and $%
dr^{\prime \mu }$ (see Eqs.(\ref{dT-1})). Analogous transformation laws
follow, of course, for the covariant components of the displacements, namely
$dr_{\nu }=g_{\mu \nu }(r)dr^{\mu }$. In view of Eqs.(\ref{T-1}) and (\ref%
{T-2}), by construction $\mathcal{J}_{\nu }^{\mu }$ and $\left( \mathcal{J}%
^{-1}\right) _{\nu }^{\mu }$ are considered respectively local functions of $%
r^{\prime }\equiv \left\{ r^{\prime \mu }\right\} $ and $r\equiv \left\{
r^{\mu }\right\} $ only and must necessarily coincide with the \emph{%
gradient forms (\ref{GR-1})-(\ref{GR-2}).} Nevertheless, since $\mathcal{J}%
_{\nu }^{\mu }$ and $\left( \mathcal{J}^{-1}\right) _{\nu }^{\mu }$ are
mutually related being inverse matrices of each other and the point
transformations are purely local, it follows that they can also both
formally be regarded as functions respectively of the variables $r^{\prime }$
and $r$.

\item \emph{LPT-requirement \#5 - }In terms of the Jacobian matrix $\mathcal{%
J}_{\nu }^{\mu }$ and its inverse $\left( \mathcal{J}^{-1}\right) _{\nu
}^{\mu }$ the fundamental \emph{LPT} $4-$\emph{tensor transformation laws}
for the group $\left\{ P\right\} $ are set for arbitrary tensors. Consider,
for example, the Riemann curvature tensor $R_{\sigma \mu \nu }^{\rho }(r)$.
In terms of an arbitrary LPT it obeys the $4-$tensor transformation law%
\begin{equation}
R_{\sigma \mu \nu }^{\rho }(r)=\mathcal{J}_{\sigma }^{\alpha }\left(
\mathcal{J}^{-1}\right) _{\alpha }^{\rho }\mathcal{J}_{\mu }^{k}\mathcal{J}%
_{\nu }^{m}R_{\beta km}^{\prime \alpha }(r^{\prime }).  \label{4-tensor}
\end{equation}

The same transformation law also requires that $4-$scalars must be left
unchanged under the action of the group $\left\{ P\right\} $.\ Thus, by
construction the $4-$scalar proper-time element $ds$, i.e., the Riemann
distance defined in terms of the equation $ds^{2}=g_{\mu \nu }(r)dr^{\mu
}dr^{\nu }\equiv g^{\mu \nu }(r)dr_{\mu }dr_{\nu }$, must satisfy the
transformation law%
\begin{equation}
ds^{2}=g_{\mu \nu }(r)dr^{\mu }dr^{\nu }=g_{\mu \nu }^{\prime }(r^{\prime
})dr^{\prime \mu }dr^{\prime \nu },  \label{scalar}
\end{equation}%
which can be equivalently expressed as
\begin{equation}
ds^{2}=g^{\mu \nu }(r)dr_{\mu }dr_{\nu }=g^{\prime \mu \nu }(r^{\prime
})dr_{\mu }^{\prime }dr_{\nu }^{\prime }.  \label{scalar-2}
\end{equation}%
Furthermore, the covariant and contravariant components of the metric
tensor, i.e., $g_{\mu \nu }(r)$ and $g^{\mu \nu }(r)$ and respectively $%
g_{\mu \nu }^{\prime }(r^{\prime })$ and $g^{\prime \mu \nu }(r^{\prime })$,
must satisfy respectively the \emph{LPT }$4-$\emph{tensor transformation laws%
}%
\begin{eqnarray}
g_{\mu \nu }^{\prime }(r^{\prime }) &=&\mathcal{J}_{\mu }^{\alpha
}(r^{\prime })\mathcal{J}_{\nu }^{\beta }(r^{\prime })g_{\alpha \beta }(r),
\label{DDT-1} \\
g^{\prime \mu \nu }(r^{\prime }) &=&\left( \mathcal{J}^{-1}(r)\right)
_{\alpha }^{\mu }\left( \mathcal{J}^{-1}(r)\right) _{\beta }^{\nu }g^{\alpha
\beta }(r),  \label{DDT-2}
\end{eqnarray}%
so that the validity of the scalar transformation laws (\ref{scalar}) and (%
\ref{scalar-2}) is warranted.

\item \emph{LPT-requirement \#6 - }Introducing the corresponding \emph{%
Lagrangian form} of the same equations, obtained by parametrizing both $%
r^{\mu }$\ and\ $r^{\prime \mu }$\ in terms of suitably-smooth time-like
world-lines $\left\{ r^{\mu }(s),s\in I\right\} $\ and\ $\left\{ r^{\prime
\mu }(s),s\in I\right\} $, Eqs.(\ref{T-1})-(\ref{T-2}) take the equivalent
form%
\begin{equation}
\left\{
\begin{tabular}{l}
$P:r^{\mu }(s)\rightarrow r^{\prime \mu }(s)=r^{\prime \mu }(r(s)),$ \\
$P^{-1}:r^{\prime \mu }(s)\rightarrow r^{\mu }(s)=r^{\mu }(r^{\prime }(s)),$%
\end{tabular}%
\right.  \label{T-1L}
\end{equation}%
whereby the displacement $4-$vectors $dr^{\mu }\equiv dr^{\mu }(s)$ and $%
dr^{\prime \mu }\equiv dr^{\prime \mu }(s)$ can be viewed as occurring
during the proper time $ds$. Then it follows that Eqs.(\ref{T-1L}) imply
also suitable transformation laws for the $4-$velocities $u^{\mu
}(s)=dr^{\mu }(s)/ds$ and $u^{\prime \mu }(s)=dr^{\prime \mu }(s)/ds$, which
by definition span the tangent space $T\mathbb{D}^{4}.$ The latter are
provided by the equations
\begin{equation}
\left\{
\begin{tabular}{l}
$u^{\mu }(s)=\mathcal{J}_{\nu }^{\mu }(r^{\prime })u^{\prime \nu }(s),$ \\
$u^{\prime \mu }(s)=\left( \mathcal{J}^{-1}\right) _{\nu }^{\mu }(r)u^{\nu
}(s).$%
\end{tabular}%
\right.  \label{TU-1}
\end{equation}%
Notice that here also the Jacobian $\mathcal{J}_{\nu }^{\mu }$ and its
inverse $\left( \mathcal{J}^{-1}\right) _{\nu }^{\mu }$\textbf{\ }must be
considered as\textbf{\ }$s-$dependent (but just only through\textbf{\ }$%
r^{\prime }=r^{\prime }(s)$\textbf{\ }and\textbf{\ }$r=r(s)$\textbf{\ }%
respectively), i.e., of the form%
\begin{eqnarray}
\mathcal{J}_{\nu }^{\mu }(r^{\prime }) &=&\mathcal{J}_{\nu }^{\mu
}(r^{\prime }(s)),  \label{JAC-1} \\
\left( \mathcal{J}^{-1}\right) _{\nu }^{\mu }(r) &=&\left( \mathcal{J}%
^{-1}\right) _{\nu }^{\mu }(r(s)).  \label{JAC-2}
\end{eqnarray}

\item \emph{LPT-requirement \#7 - }Finally, in terms of Eqs.(\ref{T-1L}) and
(\ref{TU-1}) one notices that a LPT can be formally represented in terms of
Lagrangian phase-space transformations of the type:%
\begin{equation}
\left\{
\begin{tabular}{l}
$\left\{ r^{\mu }(s),u^{\mu }(s)\right\} \rightarrow \left\{ r^{\prime \mu
}(s),u^{\prime \mu }(s)\right\} =\left\{ r^{\prime \mu }(r(s)),\left(
\mathcal{J}^{-1}\right) _{\nu }^{\mu }(r)u^{\nu }(s)\right\} ,$ \\
$\left\{ r^{\prime \mu }(s),u^{\prime \mu }(s)\right\} \rightarrow \left\{
r^{\mu }(s),u^{\mu }(s)\right\} =\left\{ r^{\mu }(r^{\prime }(s)),\mathcal{J}%
_{\nu }^{\mu }(r^{\prime })u^{\prime \nu }(s)\right\} $%
\end{tabular}%
\right.  \label{T-PHASE-1}
\end{equation}%
(\emph{LPT-phase-space transformation}), with the vectors $\left\{ r^{\mu
}(s),u^{\mu }(s)\right\} $ and $\left\{ r^{\prime \mu }(s),u^{\prime \mu
}(s)\right\} $ to be viewed as representing the phase-space states, endowed
by $4-$positions $r^{\mu }(s)$\ and $r^{\prime \mu }(s)$\ respectively, and
corresponding $4-$velocities $u^{\mu }(s)$\ and\textbf{\ }$u^{\prime \mu
}(s) $.\textbf{\ }Hence, by construction the transformation (\ref{T-PHASE-1}%
) warrants the scalar and tensor transformation laws (\ref{scalar}) and (\ref%
{DDT-1})\ and preserves the structure of the space-time $\left( \mathbf{Q}%
^{4},g\right) $.
\end{itemize}

This concludes the prescription of the LPT- functional setting required for
the validity of GCP.

It must be stressed that its adoption is of paramount importance in the
context of GR and in particular for the subsequent considerations regarding
the physical interpretations of Einstein teleparallelism. This happens at
least for the following three main motivations. The first one is that, in
validity of the \emph{LPT-requirements \#1-\#6}, and in particular the
gradient-form requirement (\ref{GR-1})-(\ref{GR-2}) for the Jacobian matrix,
Eqs.(\ref{TU-1}) are equivalent to the Eulerian equations (\ref{T-1})-\ref%
{T-2}) (and of course also to the corresponding Lagrangian equations (\ref%
{T-1L})). Hence,\emph{\ both equations actually allow one to identify
uniquely the group }$\left\{ P\right\} $ (\emph{Proposition \#1})\emph{.}

The second one concerns the very notion of particular solution to be adopted
in the context of GR for the Einstein equation. In fact, if $g_{\mu \nu
}\left( r\right) $ denotes a \emph{parametrized-solution} of the same
equation obtained with respect to a GR-frame $r^{\mu }$, the notion of
particular solution for the same equation is actually peculiar. Indeed, it
must necessarily coincide with the whole equivalence class of
parametrized-solutions, represented symbolically as $\left\{ g_{\mu \nu
}\left( r\right) \right\} $, which are mapped in each other by means of an
arbitrary LPT of the group $\left\{ P\right\} $. Such a property, which is
actually a consequence of GCP (and consequently of Classical Tensor Analysis
on Manifolds), is usually\ being referred to in GR as the so-called \emph{%
principle of frame's} (or observer's) \emph{independence} (\emph{Proposition
\#2}).

The third motivation concerns the very notion of curved space-time $(\mathbf{%
Q}^{4},g(r)),$ compared to that of the Minkowski flat space-time $(\mathbf{Q}%
^{4},\eta )$, which when expressed in orthogonal Cartesian coordinates%
\textbf{\ }$r^{\prime \mu }\equiv \left( (r^{0\prime },\left( \mathbf{r}%
^{\prime }\equiv x^{\prime },y^{\prime },z^{\prime }\right) \right) $ has
the metric tensor $\eta _{\mu \nu }=$diag$\left\{ 1,-1,-1,-1\right\} $. A
generic space-time of this type is characterized, by definition, by a
non-vanishing Riemann curvature $4-$tensor $R_{\sigma \mu \nu }^{\rho }(r)$.
As a consequence of the $4-$tensor transformation laws (\ref{DDT-1})-(\ref%
{DDT-2}) it follows that two generic space-times $(\mathbf{Q}^{4},g(r))$ and
$(\mathbf{Q}^{\prime 4},g^{\prime }(r^{\prime }))$ can be mapped in each
other by means of LPTs, and hence actually coincide, only provided the
respective metric tensors, and hence also the corresponding Riemann
curvature $4-$tensors, are transformed in each other via the same Eqs.(\ref%
{DDT-1})-(\ref{DDT-2}). Hence, it is obvious that a generic curved
space-time cannot be mapped into the said Minkowski space-time purely by
means of a LPT (\emph{Proposition} \emph{\#3}).

\section{3 - Einstein's teleparallel gravity and the teleparallel problem}

Most of the historical developments achieved so far in GR since its original
appearance in 1915 have been obtained in the framework of the GCP-setting of
GR \cite{Einst}. Nonetheless for a long time the issue has been debated
whether Relativistic Classical Mechanics and Relativistic Classical theory
of fields might exhibit in each case (possibly-different) non-local
phenomena. In the literature there are several examples of studies aimed at
extending in the context of GR the classical notions of local dynamics and
local field interactions.\ A related question is, however, whether\ there
actually exist additional non-local phenomena which might escape the
validity of GCP and require the setup of a proper theoretical framework for
their study.

As we intend to show, an instance of this type arises in the context of the
so-called teleparallel approach to GR, also known as Einstein
teleparallelism\ \cite{ein28} (see also Refs.\cite{tele1,tele2,tele3}). To
state the issue in the appropriate physical context let us briefly highlight
the basic ideas behind such an approach. This is based on the conjecture on
Einstein part that at each point $r^{\mu }$ of the space-time manifold $(%
\mathbf{Q}^{4},g(r))$ the corresponding tangent space $T\mathbb{D}^{4}$ can
be \textquotedblleft parallelized\textquotedblright . This means, in other
words, that at all $4-$positions $r^{\mu }$ $\in $ $(\mathbf{Q}^{4},g(r))$
it should be possible to cast each tangent $4-$vector $u^{\mu }(s)$ in the
form%
\begin{equation}
\left\{
\begin{tabular}{l}
$u^{\mu }(s)=M_{\nu }^{\mu }u^{\prime \nu }(s),$ \\
$u^{\prime \mu }(s)=\left( M^{-1}\right) _{\nu }^{\mu }u^{\nu }(s),$%
\end{tabular}%
\right.  \label{linear transf-0}
\end{equation}%
with $\left\{ M_{\alpha }^{\mu }\right\} $ being an invertible matrix with
inverse $\left( M^{-1}\right) _{\mu }^{\alpha }\equiv \left( M^{-1}\right)
_{\mu }^{\alpha }$.\ More precisely, according to Einstein's approach the
metric tensor of a generic curved space-time $(Q^{4},g(r))$ should satisfy
an equation in the form:%
\begin{equation}
g_{\mu \nu }\left( r\right) =\left( M^{-1}\right) _{\mu }^{\alpha }\left(
M^{-1}\right) _{\nu }^{\beta }\eta _{\alpha \beta },  \label{first equation}
\end{equation}%
or equivalent%
\begin{equation}
M_{\alpha }^{\mu }(r)M_{\beta }^{\nu }(r)g_{\mu \nu }\left( r\right) =\eta
_{\alpha \beta },  \label{last-equation}
\end{equation}%
with $\eta _{\alpha \beta }$ being here the metric tensor associated with
the flat Minkowski space-time $(\mathbf{Q}^{\prime 4}\equiv \mathbf{M}%
^{4},\eta )$ having the Lorentzian signature $(+,-,-,-)$. The goal is
therefore to determine the map%
\begin{equation}
\eta _{\alpha \beta }\leftrightarrow g_{\mu \nu }\left( r\right) ,
\end{equation}%
known as the \emph{teleparallel transformation (TT), }while Eq.(\ref{first
equation}) (or equivalent (\ref{last-equation})) will be referred to as the
\emph{TT-problem. }For definiteness, it must be stressed here what appears
to be the Einstein's key assumption underlying these equations: it is
understood in fact\textbf{\ }that in Eqs.(\ref{first equation}) and (\ref%
{last-equation}) $\eta _{\alpha \beta }$ manifestly identifies the metric
tensor of the Minkowski space-time $(M^{4},\eta )$ when expressed in terms
of orthogonal Cartesian coordinates. On the other hand it is also understood
that Eqs. (\ref{first equation}) and (\ref{last-equation}) should include
the identity transformation among their possible solutions. This means that
for consistency $g_{\mu \nu }\left( r\right) $\ can always be identified
with the metric tensor of the curved space-time $(Q^{4},g(r))$ when
expressed as a local function of the\ \emph{same Cartesian coordinates}. We
shall return on this issue in Part 2. In the present paper such a viewpoint
shall be consistently adopted in the subsequent considerations to be
developed in Section 4.

The following additional remarks\ must also be made regarding the TT-problem.

\begin{itemize}
\item The first one concerns the interpretation of Eq.(\ref{last-equation})
in the so-called tetrad formalism. It implies, in fact, that for $\mu =0,3$
the fields $M_{0}^{\mu }(r),M_{1}^{\mu }(r),M_{2}^{\mu }(r)$ and $M_{3}^{\mu
}(r)$ can simply be interpreted as a \emph{tetrad basis}, i.e., a set of
four independent real $4-$vector fields that are mutually orthogonal, i.e.,
such that for $\alpha \neq \beta :$%
\begin{equation}
e_{\alpha }^{\mu }(r)e_{\beta }^{\nu }(r)g_{\mu \nu }(r)=0.
\end{equation}%
Also, all basis $4-$vectors are unitary, in the sense that for all $\alpha
=0,3$, $\left\vert M_{\alpha }^{\mu }(r)M_{(\alpha )}^{\nu }(r)g_{\mu \nu
}(r)\right\vert =1,$ one of them ($M_{0}^{\mu }(r)$) being time-like and the
others space-like, namely%
\begin{eqnarray}
M_{0}^{\mu }(r)M_{0}^{\nu }(r)g_{\mu \nu }(r) &=&-1,  \notag \\
M_{\alpha }^{\mu }(r)M_{(\alpha )}^{\nu }(r)g_{\mu \nu }(r) &=&1,
\end{eqnarray}%
together span the 4-D tangent space at each point $r^{\mu }$ in the
space-time $(\mathbf{Q}^{4},g)$.

\item The second remark is about the choice of the curved space-time $(%
\mathbf{Q}^{4},g(r))$ in the TT-problem. It must be stressed, in fact, that
the space-time $(\mathbf{Q}^{4},g(r))$ should remain in principle arbitrary.
Therefore, it should always be possible to identify $\left( \mathbf{Q}%
^{4},g(r))\right) $ with the curved space-time having signature different
from that of the Minkowski space-time. Therefore, the solution of the
TT-problem should be possible also in the case in which $(\mathbf{Q}%
^{4},g(r))$ and $(\mathbf{M}^{4},\eta )$ have different signatures.

\item The third remark is about the ultimate goal of Einstein
teleparallelism. This emerges perspicuously from Eq.(\ref{first equation})
(or equivalent its inverse represented by Eq.(\ref{last-equation})). The
determination of the matrix $M_{\alpha }^{\mu }(r)$ solution of such an
equation will be referred to here as \emph{TT- problem}. In fact, Eq.(\ref%
{first equation}) - i.e., if a solution exists to such an equation - should
permit one to relate curved and flat space-time metric tensors,
respectively\ identified with $g_{\mu \nu }\left( r\right) $\ and $\eta
_{\alpha \beta }$.
\end{itemize}

From\textbf{\ }these premises it emerges, therefore, the fundamental problem
of establishing a map between the generic curved space-time $(\mathbf{Q}%
^{4},g)$\ indicated above and the Minkowski space-time $(\mathbf{M}^{4},\eta
)$,\ which should have a global validity, namely it should hold in the whole%
\textbf{\ }$(\mathbf{Q}^{4},g)$\textbf{\ }or at least in a finite subset of
the same space-time.\ However, such a kind of transformation cannot be
realized by means of LPT of the type (\ref{T-1})-(\ref{T-2}) in which $%
M_{\alpha }^{\mu }(r)$ is\ identified with the corresponding Jacobian (see
Eq.(\ref{GR-1}) below). This happens because the teleparallel transformation
cannot be realized by means of the group of LPT$\left\{ P\right\} $ (see\
also the related \emph{Proposition \#3} indicated above). The issue arises
whether in the context of GR the teleparallel transformation (\ref{first
equation}) (or equivalent its inverse, i.e., Eq.(\ref{last-equation})) might
actually still apply in the case of a more general type of\emph{\ non-local
point transformations} (NLPT), with the matrix $M_{\alpha }^{\mu }(r)$ to be
identified with a corresponding suitably-prescribed Jacobian matrix.

The existence of such a class of generalized GR-reference frames and
coordinate systems is actually suggested by the Einstein equivalence
principle (EEP) itself. This is expressed by two separate propositions,
which in the form presently known must both be ascribed to Albert Einstein's
1907 original formulation \cite{ein-1907} (see also Ref.\cite{ein-1911}).
The part of EEP which is mostly relevant for the current discussion is the
one usually referred to as the so-called weak equivalence principle (WEP).
This is related, in fact, to the fundamental notion of equivalence between
gravitational and inertial mass as well as to Albert Einstein's observation
that the gravitational \textquotedblleft force\textquotedblright\ as
experienced locally while standing on a massive body is actually the same as
the pseudo-force experienced by an observer in a non-inertial (accelerated)
frame of reference.\textbf{\ }Apparently there is no unique formulation of
WEP\ to be found in the literature. However, the form of WEP which is of key
importance in the following consists in the two distinct claims by Einstein
stating:\textit{\ a}) the equivalence between accelerating frames and the
occurrence of gravitational fields (see also Ref.\cite{Einst});\ \ \textit{b}%
) that \textquotedblleft \textit{local effects of motion in a curved space
(gravitation)}\textquotedblright\ should be considered\ as\
\textquotedblleft \textit{indistinguishable from those of an accelerated
observer in flat space}\textquotedblright\ \cite{ein-1907,ein-1911}.
Incidentally, it must be stressed that statement \textit{b}) is the basis of
Einstein's 1928 paper on teleparallelism.

From a historical perspective, the original introduction of WEP (and EEP) on
the part of Albert Einstein was later instrumental for the development of
GR. An interesting question\ concerns the conditions of validity of GCP and
the choice of the class of LPTs for which\ WEP applies. In fact, based on
the discussion above, the issue is whether it is possible to extend in such
a framework the class of LPTs. In particular, here we intend to look for a
more general group of point transformations, to be identified with NLPTs.
These are distinguished from the class $\left\{ P\right\} $ introduced above
and form a group of transformations denoted here as \emph{special NLPT-group
}$\left\{ P_{S}\right\} $. This new type of transformations connects two
accelerating frames, namely curvilinear coordinate systems mutually related
by means of suitable acceleration-dependent and necessarily non-local
coordinate transformations. The latter should permit one to connect globally
two suitable subsets of Lorentzian spaces which realize accessible domains
(in the sense indicated below) and are endowed with different metric tensors
having intrinsically-different Riemann tensors. Therefore, these
transformations should have the property of being globally defined and,
together with the corresponding inverse transformations, be respectively
endowed with Jacobians $M_{\alpha }^{\mu }(r)$\ and $\left( M^{-1}(r)\right)
_{\nu }^{\mu }$.

We intend to show that, provided suitable \textquotedblleft ad
hoc\textquotedblright\ restrictions are set on the class of manifolds among
which NLPTs are going to be established, a non-trivial generalization of GR
by means of the general NLPT-group $\left\{ P_{S}\right\} $. These will be
shown to be realized in terms of a suitably-prescribed diffeomorphism
between $4-$dimensional Lorentzian space-times $(\mathbf{Q}^{4},g)$\ and $(%
\mathbf{Q}^{\prime 4},g^{\prime })$ of the general form%
\begin{equation}
P_{g}:r^{\prime \mu }\rightarrow r^{\mu }=r^{\mu }\left\{ r^{\prime },\left[
r^{\prime },u^{\prime }\right] \right\} ,  \label{pg1}
\end{equation}%
with inverse transformation%
\begin{equation}
P_{g}^{-1}:r^{\mu }\rightarrow r^{\prime \mu }=r^{\prime \mu }\left\{ r,
\left[ r,u\right] \right\} .  \label{pg2}
\end{equation}%
Here the squared brackets $\left[ r^{\prime },u^{\prime }\right] $ and $%
\left[ r,u\right] $ denote possible suitable \emph{non-local dependences }in
terms of the $4-$positions $r^{\prime \mu }$,\ $r^{\mu }$\ and corresponding%
\textbf{\ }$4-$velocities $u^{\mu }\equiv \frac{dr^{\mu }}{ds}$\ and $%
u^{\prime \mu }\equiv \frac{dr^{\prime \mu }}{ds}$\ respectively. As a
consequence, Eqs.(\ref{pg1})-(\ref{pg2}) identify a new kind of point
transformations, which unlike LPTs (see Eqs.(\ref{T-1}) and (\ref{T-2}) )
are established between intrinsically different manifolds $(\mathbf{Q}%
^{4},g) $\ and $(\mathbf{Q}^{\prime 4},g^{\prime })$, i.e., which cannot be
mapped in each other purely by means of LPTs.

\section{4 - Explicit solution of the TT-problem - The NLPT-functional
setting}

Let us now pose the problem of constructing explicitly the new type of point
transformations, i.e. the NLPTs, which are involved in the representation
problem of teleparallel gravity and identifying, in the process, the
corresponding NLPT-functional setting.

For this purpose we introduce first the conjecture that, consistent with
EEP, it should be possible to generate such a transformation introducing a
suitable $4-$velocity transformation $u^{\mu }\rightarrow $\ $u^{\prime \mu
} $ which connects appropriate sets of GR-reference frames belonging to the
two space-times indicated above. Indeed, it is physically conceivable the
possibility of constructing \textquotedblleft ad hoc\textquotedblright\ $4-$%
velocity transformations which are not reducible to LPTs of the type (\ref%
{T-1}) and (\ref{T-2}).\ To show how this task can be achieved in practice,
we notice that the transformation laws for the $4-$velocity which are
realized, by assumption, by Eqs.(\ref{linear transf-0}), necessarily imply
the validity of corresponding transformation equations for the displacement $%
4-$vectors $dr^{\mu }(s)$ and $dr^{\prime \mu }(s).$ These read manifestly%
\begin{equation}
\left\{
\begin{tabular}{l}
$dr^{\mu }(s)=M_{\nu }^{\mu }dr^{\prime \nu }(s),$ \\
$dr^{\prime \mu }(s)=\left( M^{-1}\right) _{\nu }^{\mu }dr^{\nu }(s),$%
\end{tabular}%
\right.  \label{displacemtnts-NLPT}
\end{equation}%
where for generality $M_{\nu }^{\mu }$\textbf{\ }and $\left( M^{-1}\right)
_{\nu }^{\mu }$\textbf{\ }are considered of the form $M_{\nu }^{\mu }=M_{\nu
}^{\mu }(r^{\prime },r)$\textbf{\ }and $\left( M^{-1}\right) _{\nu }^{\mu
}=\left( M^{-1}\right) _{\nu }^{\mu }(r,r^{\prime })$.\textbf{\ }In analogy
with Eqs.(\ref{JAC-1}) and (\ref{JAC-2}), when evaluated along the
corresponding world-lines, it follows that they take the general functional
form%
\begin{eqnarray}
M_{\nu }^{\mu } &=&M_{\nu }^{\mu }(r^{\prime }(s),r(s)),  \label{PG-1} \\
\left( M^{-1}\right) _{\nu }^{\mu } &=&\left( M^{-1}\right) _{\nu }^{\mu
}(r(s),r^{\prime }(s)),  \label{PG-2}
\end{eqnarray}%
with $M_{\nu }^{\mu }$ and $\left( M^{-1}\right) _{\nu }^{\mu }$ being now
smooth functions of $s$ through the variables $r(s)\equiv \left\{ r^{\mu
}(s)\right\} $ and $r^{\prime }(s)\equiv \left\{ r^{\prime \mu }(s)\right\} $%
. More precisely, in analogy to the LPT-requirements recalled above, the
following prescriptions can be invoked to determine the NLPT-functional
setting:

\begin{itemize}
\item \emph{NLPT-requirement \#1 - }The coordinates $r^{\mu }$\ and\ $%
r^{\prime \mu }$\ realize by assumption physical observables and hence are
prescribed in terms of real variables, while $(\mathbf{Q}^{4},g(r))$\ and $(%
\mathbf{M}^{4},\eta )$\ must both realize $C^{k}-$differentiable Lorentzian
manifolds, with $k\geq 3.$

\item \emph{NLPT-requirement \#2} - The matrices $M_{\nu }^{\mu }$ and $%
\left( M^{-1}\right) _{\nu }^{\mu }$ are assumed to be \emph{locally}
smoothly-dependent only on $4-$position, while admitting at the same time
also possible non-local dependences. More precisely, in the case of the
Jacobian $M_{\nu }^{\mu }(r^{\prime },r)$\ the second variable $r\equiv
\left\{ r^{\mu }\right\} $\ which enters the same function can contain in
general both local and non-local implicit dependences, the former ones in
terms of $r^{\prime \mu }$. Similar considerations apply to the inverse
matrix $\left( M^{-1}\right) _{\nu }^{\mu }(r,r^{\prime })$, which besides
local explicit and implicit dependences in terms of $r^{\mu }$, may
generally include additional non-local dependences through the variable $%
r^{\prime }\equiv \left\{ r^{\prime \mu }\right\} $.

\item \emph{NLPT-requirement \#3} - The Jacobian matrix $M_{\nu }^{\mu }$
and its inverse $\left( M^{-1}\right) _{\nu }^{\mu }$ are assumed to be
generally non-gradient. In other words, at least in a subset of the two
space times $(\mathbf{M}^{4},\eta )\equiv (\mathbf{Q}^{\prime 4},g^{\prime
}) $ and $(\mathbf{Q}^{4},g)$:%
\begin{eqnarray}
M_{\nu }^{\mu }(r^{\prime },r) &\neq &\frac{\partial r^{\mu }(r^{\prime },r)%
}{\partial r^{\prime \nu }},  \label{RPG-1} \\
\left( M^{-1}\right) _{\nu }^{\mu }(r,r^{\prime }) &\neq &\frac{\partial
r^{\prime \mu }(r,r^{\prime })}{\partial r^{\nu }},  \label{RPG-2}
\end{eqnarray}%
while elsewhere they can still recover the gradient form (\ref{GR-1}) and (%
\ref{GR-2}), namely%
\begin{eqnarray}
M_{\nu }^{\mu }(r^{\prime },r) &=&\frac{\partial r^{\mu }(r^{\prime },r)}{%
\partial r^{\prime \nu }},  \label{RRPG-1} \\
\left( M^{-1}\right) _{\nu }^{\mu }(r,r^{\prime }) &=&\frac{\partial
r^{\prime \mu }(r,r^{\prime })}{\partial r^{\nu }}.  \label{RRPG-2}
\end{eqnarray}%
In both cases the partial derivative are performed with respect to the local
dependences only.

\item \emph{NLPT-requirement \#4 -} Introducing the (proper-time) line
elements $ds$, $ds^{\prime }$ in the two space-times $(\mathbf{Q}^{4},g)$
and $(\mathbf{M}^{4},\eta )\equiv (\mathbf{Q}^{\prime 4},g^{\prime })$
defined respectively according to Eq.(\ref{scalar-2}) and so that%
\begin{eqnarray}
ds^{2} &=&g_{\mu \nu }(r)dr^{\mu }dr^{\nu }, \\
ds^{\prime 2} &=&g_{\mu \nu }^{\prime }(r^{\prime })dr^{\prime \mu
}dr^{\prime \nu }\equiv \eta _{\mu \nu }dr^{\prime \mu }dr^{\prime \nu },
\end{eqnarray}
the \emph{isometric constraint condition}%
\begin{equation}
ds=ds^{\prime }
\end{equation}%
is set. This implies that the equation
\begin{equation}
g_{\mu \nu }(r)dr^{\mu }dr^{\nu }=\eta _{\mu \nu }dr^{\prime \mu }dr^{\prime
\nu }  \label{IMPLICATION NLPT-4}
\end{equation}%
must hold.

\item \emph{NLPT-requirement \#5 - }Finally, we shall assume that the $4-$%
positions $r^{\mu }(s)$ and $r^{\prime \mu }(s)$ spanning the corresponding
space-times $(\mathbf{Q}^{4},g)$ and $(\mathbf{M}^{4},\eta )$ are
represented in terms of the same Cartesian coordinates, i.e.,%
\begin{equation}
r^{\mu }\equiv \left\{ ct,\left( \mathbf{r}\equiv x,y,z\right) \right\}
\label{Cartesian-1}
\end{equation}%
and%
\begin{equation}
r^{\prime \mu }\equiv \left\{ ct^{\prime },\left( \mathbf{r}^{\prime }\equiv
x^{\prime },y^{\prime },z^{\prime }\right) \right\} .  \label{Cartesian-2}
\end{equation}
\end{itemize}

Let us now briefly analyze the implications of these Requirements. First,
Eqs.(\ref{displacemtnts-NLPT}) (or equivalent Eqs.(\ref{linear transf-0}))
can be integrated at once performing the integration along suitably-smooth
time- (or space-) like world lines $r^{\mu }(s)$\ and\textbf{\ }$r^{\prime
\mu }(s)$%
\begin{equation}
\left\{
\begin{array}{c}
P_{S}:r^{\mu }(s)=r^{\mu }(s_{o})+\int_{s_{o}}^{s}d\overline{s}M_{\nu }^{\mu
}(r^{\prime },r)u^{\prime \nu }(\overline{s}), \\
P_{S}^{-1}:r^{\prime \mu }(s)=r^{\prime \mu }(s_{o})+\int_{s_{o}}^{s}d%
\overline{s}\left( M^{-1}\right) _{\nu }^{\mu }(r,r^{\prime })u^{\nu }(%
\overline{s}),%
\end{array}%
\right.  \label{NLPT-1}
\end{equation}%
where the initial condition is set%
\begin{equation}
r^{\mu }(s_{o})=r^{\prime \mu }\left( s_{o}\right) .
\label{initisal condition}
\end{equation}%
Transformations (\ref{NLPT-1}) will be referred to as \emph{special NLPT} in
\emph{Lagrangian} \emph{form, }the family of such transformations
identifying the special NLPT-group\emph{\ }$\left\{ P_{S}\right\} $, i.e., a
suitable subset of the group of general NLPT-group $\left\{ P_{g}\right\} $%
\emph{.} The subsets of two space-times $(\mathbf{Q}^{4},g)$ and $(\mathbf{Q}%
^{\prime 4},g^{\prime })\equiv (\mathbf{M}^{4},\eta )$ which are mapped in
each other by a special NLPT, both assumed to have non-vanishing measure,
will be referred to as \emph{accessible sub-domains}. Depending on the
signature of $(\mathbf{Q}^{4},g)$ an accessible subset of the same
space-time can be covered in principle either by time- (or space-) like
world-lines.

Notice that the Jacobians $M_{\nu }^{\mu }(r^{\prime },r)$ and $\left(
M^{-1}\right) _{\nu }^{\mu }(r,r^{\prime })$ remain still in principle
arbitrary. In particular, in case they take the gradient forms (\ref{RRPG-1}%
) and (\ref{RRPG-2}) the Lagrangian LPT defined by Eqs.(\ref{T-1L}) is
manifestly recovered. Furthermore, Eqs.(\ref{linear transf-0}), or
equivalent Eqs.(\ref{NLPT-1}), can be also represented in terms of the
equations for the infinitesimal $4-$displacements, given by Eq.(\ref%
{displacemtnts-NLPT}). In particular, assuming the matrix $M_{\nu }^{\mu }$
to be continuously connected to the identity $\delta _{\nu }^{\mu }$,
implies that the Jacobian matrix $M_{\nu }^{\mu }$ and its inverse $\left(
M^{-1}\right) _{\nu }^{\mu }$ can always be represented in the form%
\begin{eqnarray}
M_{\nu }^{\mu } &=&\delta _{\nu }^{\mu }+\mathcal{A}_{\nu }^{\mu
}(r,r^{\prime }),  \label{PPG-1} \\
\left( M^{-1}\right) _{\nu }^{\mu } &=&\delta _{\nu }^{\mu }+\mathcal{B}%
_{\nu }^{\mu }(r,r^{\prime }),  \label{PPG-2}
\end{eqnarray}%
with $\mathcal{A}_{\nu }^{\mu }$\textbf{\ }and\textbf{\ }$\mathcal{B}_{\nu
}^{\mu }$ being suitable \emph{transformation matrices},\ which are\
mutually related by matrix inversion. Hence, in terms of Eqs.(\ref{PPG-1})-(%
\ref{PPG-2}), the special NLPT in Lagrangian form (\ref{NLPT-1}) yields then
the corresponding Lagrangian and Eulerian forms:%
\begin{equation}
\left\{
\begin{array}{c}
r^{\mu }(s)=r^{\prime \mu }(s)+\int_{s_{o}}^{s}d\overline{s}\mathcal{A}_{\nu
}^{\mu }(r^{\prime },r)u^{\prime \nu }(\overline{s}) \\
r^{\prime \mu }(s)=r^{\mu }(s)+\int_{s_{o}}^{s}d\overline{s}\mathcal{B}_{\nu
}^{\mu }(r^{\prime },r)u^{\nu }(\overline{s})%
\end{array}%
\right. ,  \label{special-NLPT-1}
\end{equation}%
\begin{equation}
\left\{
\begin{array}{c}
r^{\mu }=r^{\mu }+\int_{r^{\prime \nu }(s_{o})}^{r^{\prime \nu }}dr^{\prime
\nu }\mathcal{A}_{\nu }^{\mu }(r^{\prime },r) \\
r^{\prime \mu }=r^{\mu }+\int_{r^{\nu }(s_{o})}^{r^{\nu }}dr^{\nu }\mathcal{B%
}_{\nu }^{\mu }(r,r^{\prime })%
\end{array}%
\right. .  \label{special-NLPT-2}
\end{equation}%
We stress that, in difference with the treatment of LPT, in the proper-time
integral on the rhs of Eqs.(\ref{NLPT-1}) and (\ref{special-NLPT-1}) the
tangent-space curve\textbf{\ }$u^{\prime \nu }(\overline{s})$\ (respectively
$u^{\nu }(\overline{s})$) must be considered as an independent variable.
This is a peculiar feature of Eqs.(\ref{NLPT-1}) which cannot be avoided.
The reason lies in the fact that there is no way by which $u^{\prime \nu }(%
\overline{s})$\ (and $u^{\nu }(\overline{s})$)\ can be uniquely prescribed
by means of the same equations. Indeed, equations (\ref{NLPT-1}) (or
equivalent (\ref{special-NLPT-1}) and (\ref{special-NLPT-2})) together with
Eqs.(\ref{linear transf-0}) truly establish a phase-space transformation of
the form:%
\begin{equation}
\left\{
\begin{tabular}{l}
$\left\{ r^{\mu }(s),u^{\mu }(s)\right\} \rightarrow \left\{ r^{\prime \mu
}(s),u^{\prime \mu }(s)\right\} =\left\{ r^{\prime \mu }\left\{ r(s),\left[
r,u\right] \right\} ,\left( \mathcal{M}^{-1}\right) _{\nu }^{\mu }(r)u^{\nu
}(s)\right\} $ \\
$\left\{ r^{\prime \mu }(s),u^{\prime \mu }(s)\right\} \rightarrow \left\{
r^{\mu }(s),u^{\mu }(s)\right\} =\left\{ r^{\mu }\left\{ r^{\prime }(s),%
\left[ r^{\prime },u^{\prime }\right] \right\} ,\mathcal{M}_{\nu }^{\mu
}(r^{\prime })u^{\prime \nu }(s)\right\} $%
\end{tabular}%
\right. .  \label{T-PHASE-2}
\end{equation}%
This will be referred to as\ \emph{NLPT-phase-space transformation. }The
latter apply to a new type of reference frame, denoted as\emph{\ extended
GR-frames}, which are represented by the vectors $\left\{ r^{\mu }(s),u^{\mu
}(s)\right\} $ and $\left\{ r^{\prime \mu }(s),u^{\prime \mu }(s)\right\} $
respectively. These can be viewed as phase-space states (of the
corresponding extended GR-frames) having respectively\emph{\ }$4-$positions $%
r^{\mu }(s)$ and $r^{\prime \mu }(s)$ and $4-$velocities $u^{\mu }(s)$ and $%
u^{\prime \mu }(s).$\textbf{\ }Finally, let us mention that the
transformation (\ref{T-PHASE-2}), in contrast with (\ref{T-PHASE-1}),
obviously does not preserve the structure of the space-times $\left( \mathbf{%
Q}^{4},g\right) $ and$.\left( \mathbf{M},\eta \right) .$ Nevertheless the
scalar transformation law (\ref{scalar}) is still by construction warranted,
while at the same time the metric tensor satisfies by construction the
TT-problem, i.e., Eq.(\ref{first equation}).

Let us now show how the matrices $A_{\nu }^{\mu }$\ and $B_{\nu }^{\mu }$\
can be explicitly determined in terms of the teleparallel transformation (%
\ref{first equation}). The relevant results, which actually prescribe the
general form of related NLPT, are summarized by the following proposition.

\bigskip

\textbf{THM.1 - Realization of the special NLPT-group} $\left\{
P_{S}\right\} $ \textbf{for the TT-problem}

\emph{Let us assume that }$(\mathbf{Q}^{4},g)$ \emph{and }$(\mathbf{Q}%
^{\prime 4},g^{\prime })\equiv (\mathbf{M}^{\prime 4},\eta )$ \emph{identify
respectively a generic curved space-time and the Minkowski space-time both
parametrized in terms of} \emph{orthogonal Cartesian coordinates (\ref%
{algebr-eq-ex}) and (\ref{Cartesian-2}).}

\emph{Then, given validity of the NLPT-Requirement \#1-\#5, the following
propositions hold.}

P$_{1})$ \emph{In the accessible sub-domain of} $(\mathbf{Q}^{4},g)$\emph{\
the teleparallel transformation (\ref{first equation}) (or equivalent its
inverse, i.e., Eq.(\ref{last-equation})), relating }$(\mathbf{Q}^{4},g)$
\emph{with the Minkowski space-time }$(\mathbf{Q}^{\prime 4},g^{\prime
})\equiv (\mathbf{M}^{\prime 4},\eta ),$ \emph{is realized by a non-local
point transformation of the type (\ref{NLPT-1}), or equivalent (\ref%
{special-NLPT-1}) and (\ref{special-NLPT-2}), with a Jacobian }$M_{\mu
}^{\nu }$ \emph{and its inverse} $\left( M^{-1}\right) _{\mu }^{\alpha }$
\emph{being of the form (\ref{PG-1}) and (\ref{PG-2}) respectively. This is
required to satisfy the }$\emph{4-}$\emph{tensor transformation law
prescribed by the matrix equations}%
\begin{equation}
g_{\mu \nu }\left( r\right) =\left( M^{-1}\right) _{\mu }^{\alpha
}(r,r^{\prime })\left( M^{-1}\right) _{\nu }^{\beta }(r,r^{\prime })\eta
_{\alpha \beta },  \label{transf-2}
\end{equation}%
\emph{and similarly its inverse (see Eq.(\ref{last-equation})) where }$%
g_{\mu \nu }\left( r\right) $\emph{\ identifies a prescribed symmetric
metric tensor associated with the space-time }$(\mathbf{Q}^{4},g)$,\emph{\
by assumption expressed in the Cartesian coordinates (\ref{Cartesian-1}).}
\emph{Hence, }$\left( M^{-1}\right) _{\mu }^{\alpha }(r,r^{\prime })$ \emph{%
necessarily coincides with the} \emph{Jacobian matrix of the TT-problem (see
Eq.\ref{first equation})).}

P$_{2})$ \emph{The set of special NLPT has the structure of a group.}

\emph{Proof -} Let us prove proposition P$_{1}$. For this purpose it is
sufficient to construct explicitly a possible, i.e., non-unique, realization
of the NLPT and the corresponding set $\left\{ P_{S}\right\} $, satisfying
Eq.(\ref{transf-2}). In fact, let us consider the equation for the
infinitesimal\emph{\ }$4-$displacement $dr^{\prime \mu }$ (see Eq.(\ref%
{displacemtnts-NLPT})), which in validity of Eq.(\ref{PPG-2}) becomes%
\begin{equation}
dr^{\prime \mu }=\left[ \delta _{\nu }^{\mu }+\mathcal{B}_{\nu }^{\mu
}(r,r^{\prime })\right] dr^{\nu },  \label{infinitesima-NLPT-1}
\end{equation}%
and similarly%
\begin{equation}
dr^{\mu }=\left[ \delta _{\nu }^{\mu }+\mathcal{A}_{\nu }^{\mu }(r,r^{\prime
})\right] dr^{\prime \nu },  \label{infinitesima-NLPT-2}
\end{equation}%
where the matrices $\mathcal{B}_{\nu }^{\mu }(r,r^{\prime })$ and $\mathcal{A%
}_{\nu }^{\mu }(r,r^{\prime })$ are suitably related. Substituting $%
dr^{\prime \mu }$ on the rhs of the last equation and invoking the
independence of the components of the infinitesimal displacement $dr^{\mu }$%
, this means for consistency that the covariant components of the metric
tensor, i.e., $g_{\mu \nu }(r)$ and respectively $g_{\mu \nu }^{\prime
}(r^{\prime })\equiv \eta _{\mu \nu }$ must satisfy the $4-$tensor
transformation law (\ref{transf-2}). Such a tensor equation delivers,
therefore, a set of 10 algebraic equations.\ Their solution can be
determined in a straightforward way for the 16 components of the matrix $%
\mathcal{B}_{\nu }^{\mu }(r,r^{\prime })$. For example, one of these
equations reads%
\begin{equation}
g_{00}\left( r\right) =\left[ 1+\mathcal{B}_{0}^{0}(r,r^{\prime })\right]
^{2}-\left[ \mathcal{B}_{0}^{1}(r,r^{\prime })\right] ^{2}-\left[ \mathcal{B}%
_{0}^{2}(r,r^{\prime })\right] ^{2}-\left[ \mathcal{B}_{0}^{3}(r,r^{\prime })%
\right] ^{2}.  \label{algebr-eq-ex}
\end{equation}%
The remaining equations following from Eq.(\ref{transf-2}) are not reported
here for brevity.

One can nevertheless show that the solution to this set is non-unique. In
fact, due to the freedom in the choice of the matrix elements of $\mathcal{B}%
_{\nu }^{\mu }(r,r^{\prime })$, the latter can in principle be chosen
arbitrarily by suitably prescribing 6 components of the same matrix. A
particular solution is obtained, for example, by requiring validity of the
constraint equations%
\begin{eqnarray}
\mathcal{B}_{3}^{0}(r,r^{\prime }) &=&\mathcal{B}_{3}^{1}(r,r^{\prime })=%
\mathcal{B}_{0}^{1}(r,r^{\prime })=0,  \notag \\
\mathcal{B}_{0}^{2}(r,r^{\prime }) &=&\mathcal{B}_{1}^{2}(r,r^{\prime })=%
\mathcal{B}_{3}^{2}(r,r^{\prime })=0.
\end{eqnarray}%
The surviving components of $\mathcal{B}_{\nu }^{\mu }$ are then determined
by the same algebraic equations of the set (\ref{transf-2}). From these
considerations it follows that necessarily it must be $\mathcal{B}_{\nu
}^{\mu }=\mathcal{B}_{\nu }^{\mu }(r)$. In particular, here we notice that
all diagonal components $\mathcal{B}_{i}^{i}(r)$ for $i=0,3$ can be viewed
as determined, up to an arbitrary sign, by the diagonal components of the
metric tensor $g_{\mu \mu }\left( r\right) $. Instead, the remaining
non-diagonal matrix elements are then prescribed in terms of the
non-diagonal components of the metric tensor, which follow analogously from
the corresponding 6 equations of the set. Then, both the $4-$displacement
transformations (\ref{infinitesima-NLPT-1})\ and their inverse (\ref%
{infinitesima-NLPT-2})\ ones exist and can be non-uniquely prescribed. An
example of possible realization is given by%
\begin{equation}
\left\{
\begin{array}{c}
dr^{\prime 0}=\left[ 1+\mathcal{B}_{0}^{0}\right] dr^{0}+\mathcal{B}%
_{2}^{0}dr^{2} \\
dr^{\prime 1}=\left[ 1+\mathcal{B}_{1}^{1}\right] dr^{1}+\mathcal{B}%
_{0}^{1}dr^{0}+\mathcal{B}_{2}^{1}dr^{2} \\
dr^{\prime 2}=\left[ 1+\mathcal{B}_{2}^{2}\right] dr^{2} \\
dr^{\prime 3}=\left[ 1+\mathcal{B}_{3}^{3}\right] dr^{3}+\mathcal{B}%
_{2}^{3}dr^{2}%
\end{array}%
\right. ,  \label{inf-1a}
\end{equation}%
with determinant%
\begin{equation}
\left\vert
\begin{array}{cccc}
1+\mathcal{B}_{0}^{0} & 0 & \mathcal{B}_{2}^{0} & 0 \\
\mathcal{B}_{0}^{1} & 1+\mathcal{B}_{1}^{1} & \mathcal{B}_{2}^{1} & 0 \\
0 & 0 & \left[ 1+\mathcal{B}_{2}^{2}\right] & 0 \\
0 & 0 & \mathcal{B}_{2}^{3} & \left[ 1+\mathcal{B}_{3}^{3}\right]%
\end{array}%
\right\vert =\prod\limits_{i=0,3}\left( 1+\mathcal{B}_{i}^{i}\right) ,
\end{equation}%
to be assumed as non-vanishing, and with inverse transformation%
\begin{equation}
\left\{
\begin{array}{c}
dr^{0}=\frac{1}{1+\mathcal{B}_{0}^{0}}\left[ dr^{\prime 0}+\frac{\mathcal{B}%
_{2}^{0}}{\frac{1}{1+\mathcal{B}_{2}^{2}}}dr^{\prime 2}\right] \\
dr^{1}=\frac{1}{1+\mathcal{B}_{1}^{1}}\left[ dr^{\prime 1}-\frac{\mathcal{B}%
_{0}^{1}}{1+\mathcal{B}_{0}^{0}}dr^{\prime 0}-\frac{1}{1+\mathcal{B}_{2}^{2}}%
\frac{\mathcal{B}_{0}^{1}\mathcal{B}_{2}^{0}}{1+\mathcal{B}_{0}^{0}}%
dr^{\prime 2}\right] \\
dr^{2}=\frac{1}{1+\mathcal{B}_{2}^{2}}dr^{\prime 2} \\
dr^{3}=\frac{1}{1+\mathcal{B}_{3}^{3}}\left[ dr^{\prime 3}-\frac{\mathcal{B}%
_{2}^{3}}{1+\mathcal{B}_{2}^{2}}dr^{\prime 2}\right]%
\end{array}%
.\right.  \label{inf-2}
\end{equation}%
In particular, from Eqs.(\ref{inf-2}) one can easily evaluate in terms of $%
\mathcal{B}_{\nu }^{\mu }(r)$\ the precise expression taken by of the matrix
$\mathcal{A}_{\nu }^{\mu }$\ which appears in Eqs.(\ref{inf-1}). Hence one
finds that necessarily $\mathcal{A}_{\nu }^{\mu }=\mathcal{A}_{\nu }^{\mu
}\left( r\right) $, with $r\equiv \left\{ r^{\mu }\right\} $ being now
considered as prescribed by means of the NLPTs (\ref{special-NLPT-1}).
Finally, the corresponding finite NLPTs generated by Eqs.(\ref{inf-1}) and (%
\ref{inf-2}) can always be equivalently represented in terms Eqs.(\ref%
{NLPT-1}).

Next, let us prove proposition P$_{2}.$\ For this purpose we first notice
that the Jacobian $\mathcal{J}_{\nu }^{\mu }\equiv \delta _{\nu }^{\mu }+%
\mathcal{A}_{\nu }^{\mu }(r,r^{\prime })$\ admits the inverse which by
construction coincides with $\left( \mathcal{J}^{-1}\right) _{\rho }^{\mu
}\equiv \delta _{\nu }^{\mu }+\mathcal{B}_{\nu }^{\mu }(r^{\prime },r)$.\
Furthermore, let us consider two special NLPTs%
\begin{equation}
\mathcal{J}_{(i)\nu }^{\mu }\equiv \delta _{\nu }^{\mu }+\mathcal{A}_{(i)\nu
}^{\mu }(r_{(i)},r^{\prime })  \label{Jacobinas-(i)}
\end{equation}%
which map the space-times \emph{\ }$(\mathbf{Q}_{(i)}^{4},g)$ $($for $i=1,2)$
onto $(\mathbf{M}^{4},\eta )$. Requiring that both the corresponding
admissible subsets of\textbf{\ }$(\mathbf{M}^{4},\eta )$\textbf{\ }and their
intersection have a non-vanishing measure, the product of two special NLPT
is defined on such a set. Its Jacobian is%
\begin{equation}
\mathcal{J}_{\nu }^{\mu }=\left( \delta _{\alpha }^{\mu }+\mathcal{A}%
_{(1)\alpha }^{\mu }(r_{(1)},r^{\prime })\right) \left( \delta _{\nu
}^{\alpha }+\mathcal{B}_{(2)\nu }^{\alpha }(r^{\prime },r_{(2)})\right)
=\left( \delta _{\nu }^{\mu }+\mathcal{C}_{\nu }^{\mu }(r_{(1)},r^{\prime
},r_{(2)})\right) ,  \label{Prtoduct-(i)}
\end{equation}%
with $\mathcal{C}_{\nu }^{\mu }(r_{(1)},r^{\prime },r_{(2)})\equiv \mathcal{A%
}_{(1)\nu }^{\mu }(r_{(1)},r^{\prime })+\mathcal{B}_{(2)\nu }^{\mu
}(r^{\prime },r_{(2)})+\mathcal{A}_{(1)\alpha }^{\mu }(r_{(1)},r^{\prime })%
\mathcal{B}_{(2)\nu }^{\alpha }(r^{\prime },r_{(2)})$. It follows that in
such a circumstance the product of the two special NLPT belongs necessarily
to the same set $\left\{ P_{S}\right\} ,$\ which is therefore a group.

\textbf{Q.E.D.}

\bigskip

THM.1 provides the formal solution of the Einstein's TT-problem in the
framework of the theory of NLPT. This is achieved by means of the
introduction of a non-local phase-space transformation of the type (\ref%
{T-PHASE-1}), which is realized by means of a special NLPT (\ref{NLPT-1})
and the corresponding 4-velocity transformation law (\ref{linear transf-0}).
In this reference the following comments must be mentioned.

\begin{itemize}
\item First, the NLPT-functional setting has been prescribed in terms of the%
\textbf{\ }special NLPT-group $\left\{ P_{S}\right\} ,$\textbf{\ }determined
here by Eqs.(\ref{NLPT-1}) together with the\textbf{\ }\emph{%
NLPT-Requirements \#1-\#4.}

\item Due to the non-uniqueness of the matrix $\mathcal{B}_{\nu }^{\mu }(r)$
solution of the TT- problem (see Eq.(\ref{transf-2})), and of the related
matrix $\mathcal{A}_{\nu }^{\mu }$, the realization of the NLPT
transformation (\ref{inf-1}) [and hence (\ref{inf-2})] yielding the solution
of the TT-problem is manifestly non-unique too. For a prescribed curved
space-time $(\mathbf{Q}^{4},g)$ which is parametrized in terms of the
Cartesian coordinates, the ensemble of NLPTs which provide particular
solutions of the TT-problem will be denoted as $\left\{ P_{g}\right\} _{TT}$.

\item Both for Eqs.\ref{inf-1a}) and (\ref{inf-2}) the corresponding
Jacobians determined by means of Eqs.(\ref{PPG-1}) and (\ref{PPG-2}) take by
construction, and consistent with Eqs.(\ref{RPG-1})-(\ref{RPG-2}), a
manifest \emph{non-gradient form.} This follows immediately from \emph{%
Proposition \#1 }thanks to the validity of Eq.(\ref{transf-2}) and the
requirement that $(\mathbf{Q}^{4},g)$ is a curved space-time.

\item In terms of the Jacobian matrix $M_{\nu }^{\mu }(r^{\prime },r)$ (and
its inverse $\left( M^{-1}\right) _{\nu }^{\mu }(r,r^{\prime })$)\textbf{\ }%
Eq.(\ref{transf-2}) means that $g_{\mu \nu }(r)$ should actually satisfy the
original Einstein's equations (\ref{first equation}) and (\ref{last-equation}%
). The latter can be interpreted as $4$\emph{-tensor transformation laws }%
for the matrix tensor $g_{\mu \nu }(r)$\emph{.}

\item Similarly and in analogy with Eq. (\ref{scalar}) holding in the case
of LPT, the validity of the scalar transformation law (\ref{scalar-2}) is
warranted also in the case of NLPT, thanks to the transformation law (\ref%
{transf-2}).

\item Finally, the transformation law (\ref{transf-2}) for the metric tensor
can be interpreted as tensor transformation law with respect to the special
NLPT-group\emph{\ }$\left\{ P_{S}\right\} $\emph{.} This will be referred to
as \emph{NLPL }$4-$\emph{tensor transformation law}. In terms of the same
Jacobian matrix\textbf{\ }$M_{\nu }^{\mu }(r^{\prime },r)$\textbf{\ }and its
inverse $\left( M^{-1}\right) _{\nu }^{\mu }(r,r^{\prime })$, analogous NLPT
$4-$tensor transformation laws can be set in principle for tensors of
arbitrary order.\textbf{\ }Nevertheless, it must be noted that -
specifically because of the validity of the same transformation law (\ref%
{transf-2}) - such a type of tensor transformation laws cannot be fulfilled
by the Riemann curvature tensor $R_{\sigma \mu \nu }^{\rho }(r)$, the reason
being that it manifestly vanishes identically in the\ case of the Minkowski
space-time.
\end{itemize}

\section{5 - Conditions of existence of special NLPT for the TT-problem}

A fundamental aspect of the theory developed here concerns the conditions of
existence of the family of special NLPT determined according to THM.1. In
this regard we notice that the identification of the physical domain of
existence involves the (possibly non-unique) prescription of the actual
possible realization of the NLPT and of the corresponding subset of $(%
\mathbf{Q}^{4},g)$ which can be mapped onto the Minkowski space-time $(%
\mathbf{Q}^{\prime 4},g^{\prime })\equiv (\mathbf{M}^{\prime 4},\eta )$. It
is obvious that NLPT, just like LPT, can only be defined in the accessible
sub-domains of $(\mathbf{Q}^{4},g)$, namely the connected subsets which in
the curved space-time can be covered by time- (or space-) like world-lines $%
r^{\mu }(s)$ which are endowed with a finite $4-$velocity. Nevertheless, the
components of the same $4-$velocity can still be in principle
arbitrarily-large, so that the corresponding world-line can be arbitrarily
close to light trajectories (and therefore to the light cones).

Another aspect of the existence problem for NLPTs is related to the
solubility conditions of the algebraic equations arising in THM.1, which
follow from the requirement that all components of the matrix $\mathcal{B}%
_{\nu }^{\mu }(r,r^{\prime })$ should be real. For example, in the case of
Eq.(\ref{algebr-eq-ex}) the corresponding condition is determined by the
inequality%
\begin{equation}
g_{00}\left( r\right) +\left[ \mathcal{B}_{0}^{1}(r,r^{\prime })\right] ^{2}+%
\left[ \mathcal{B}_{0}^{2}(r,r^{\prime })\right] ^{2}+\left[ \mathcal{B}%
_{0}^{3}(r,r^{\prime })\right] ^{2}\geq 0  \label{inequalities-THM-1}
\end{equation}%
It must be stressed that the validity of inequalities of this type for the
remaining equations in general cannot be warranted in the whole admissible
subset of the space-time $(\mathbf{Q}^{4},g)$, i.e., in particular in the
subset in which $ds^{2}>0$.\ On the other hand, \textquotedblleft \textit{a
priori}\textquotedblright\ the symmetric metric tensor $g_{\mu \nu }\left(
r\right) $\ must be regarded in principle as completely arbitrary. Hence it
is obvious that such inequalities following from THM.1 cannot place any
\textquotedblleft unreasonable\textquotedblright\ physical constraint on the
same tensor $g_{\mu \nu }\left( r\right) $.

In fact, consider the case in which the metric tensor $g_{\mu \nu }(r)$\ has
the signature $(+,-,-,-)$ and is also diagonal, namely $g_{\mu \nu
}(r)=diag\left\{ g_{00}\left( r\right) ,g_{11}\left( r\right) ,g_{22}\left(
r\right) ,g_{33}\left( r\right) \right\} $.\ Then, necessarily the metric
tensor must be such that everywhere in the same admissible subset $%
g_{00}\left( r\right) >0$,\ while $g_{11}\left( r\right) ,g_{22}\left(
r\right) ,g_{33}\left( r\right) <0$.\ As a consequence the functional class $%
\left\{ P_{g}\right\} _{TT}$\ contains transformations which may not exist
everywhere in the same set. In fact, some of the inequalities of the group (%
\ref{inequalities-THM-1}) which involve the spatial components, i.e., $%
g_{ii}\left( r\right) $\ (with $i=1,2,3$), must be considered as local,
i.e., are subject to the condition of local validity of the same
inequalities. Although NLPT of this kind are physically admissible, the
question arises whether particular solutions actually exist which are not
required to fulfill the same inequalities (\ref{inequalities-THM-1}). These
solutions, if they actually exist, have therefore necessarily a global
character, i.e., they are defined everywhere in the same admissible subset
of $(\mathbf{Q}^{4},g)$. In view of these considerations, since the only
acceptable physical restriction on $g_{\mu \nu }\left( r\right) $\ concerns
its signature, it can be shown that global validity is warranted everywhere
in\ $(\mathbf{Q}^{4},g)$\ provided the following two sets of constraints are
required to hold:%
\begin{equation}
\left[ \mathcal{B}_{0}^{1}(r,r^{\prime })\right] ^{2}+\left[ \mathcal{B}%
_{0}^{2}(r,r^{\prime })\right] ^{2}+\left[ \mathcal{B}_{0}^{3}(r,r^{\prime })%
\right] ^{2}=0,  \label{FIRST CONSTRAINT-THM-1}
\end{equation}%
and%
\begin{eqnarray}
\left[ \mathcal{B}_{1}^{0}(r,r^{\prime })\right] ^{2}-\left[ \mathcal{B}%
_{1}^{2}(r,r^{\prime })\right] ^{2}-\left[ \mathcal{B}_{1}^{3}(r,r^{\prime })%
\right] ^{2} &\geq &-\inf \left\{ g_{11}\left( r\right) \right\} ,  \notag \\
\left[ \mathcal{B}_{2}^{0}(r,r^{\prime })\right] ^{2}-\left[ \mathcal{B}%
_{2}^{1}(r,r^{\prime })\right] ^{2}-\left[ \mathcal{B}_{2}^{3}(r,r^{\prime })%
\right] ^{2} &\geq &-\inf \left\{ g_{22}\left( r\right) \right\} ,  \notag \\
\left[ \mathcal{B}_{3}^{0}(r,r^{\prime })\right] ^{2}-\left[ \mathcal{B}%
_{3}^{1}(r,r^{\prime })\right] ^{2}-\left[ \mathcal{B}_{3}^{2}(r,r^{\prime })%
\right] ^{2} &\geq &-\inf \left\{ g_{33}\left( r\right) \right\} .
\label{SECOND  CONSTRAINT-THM-1}
\end{eqnarray}%
The first equations actually requires the following $3$ independent
equations
\begin{equation}
\mathcal{B}_{0}^{1}(r,r^{\prime })=\mathcal{B}_{0}^{2}(r,r^{\prime })=%
\mathcal{B}_{0}^{3}(r,r^{\prime })=0  \label{CONSTRAINTS-THM-1-B}
\end{equation}%
to apply separately. Particular solutions of the components of $\mathcal{B}%
_{\nu }^{\mu }$ satisfying the $3$ constraint equations (\ref%
{CONSTRAINTS-THM-1-B}) and either the $3$ inequalities (\ref{SECOND
CONSTRAINT-THM-1}), or corresponding equations obtained replacing the
inequality symbol with $=,$ will be denoted respectively as\emph{\ partially}
\emph{unconditional }or \emph{unconditional. }In both cases it is immediate
to show that these solutions are non-unique, even if in all cases the
transformation matrix is again a local function of $r$, i.e., $\mathcal{B}%
_{\nu }^{\mu }=\mathcal{B}_{\nu }^{\mu }(r)$.

In particular, here we notice that all the diagonal components $B_{i}^{i}(r)$%
\ for $0=1,3$\ can be viewed as determined, up to an arbitrary sign, by the
diagonal components of the metric tensor $g_{\mu \mu }\left( r\right) $.\
Instead, the remaining non-diagonal matrix elements are then prescribed in
terms of the non-diagonal components of the metric tensor, which follow
analogously from the set of equations mentioned in THM.1. In validity of the
constraints given above, i.e., both for partially unconditional or
unconditional particular solutions, the $4-$displacement transformations (%
\ref{infinitesima-NLPT-1})\ become%
\begin{equation}
\left\{
\begin{array}{c}
dr^{\prime 0}=\left[ 1+\mathcal{B}_{0}^{0}\right] dr^{0}+\mathcal{B}%
_{1}^{0}dr^{1}+\mathcal{B}_{2}^{0}dr^{2}+\mathcal{B}_{3}^{0}dr^{3} \\
dr^{\prime 1}=\left[ 1+\mathcal{B}_{1}^{1}\right] dr^{1}+\mathcal{B}%
_{2}^{1}dr^{2}+\mathcal{B}_{3}^{1}dr^{3} \\
dr^{\prime 2}=\left[ 1+\mathcal{B}_{2}^{2}\right] dr^{2}+dr^{1}\mathcal{B}%
_{1}^{2}+\mathcal{B}_{3}^{2}dr^{3} \\
dr^{\prime 3}=\left[ 1+\mathcal{B}_{3}^{3}\right] dr^{3}+\mathcal{B}%
_{1}^{3}dr^{1}+\mathcal{B}_{2}^{3}dr^{2}%
\end{array}%
\right. .  \label{inf-1}
\end{equation}%
Similarly, one can show that also the corresponding inverse NLPT exist.

Another relevant issue concerns the role of the non-local dependences
entering the integral equations (\ref{special-NLPT-1}) and (\ref%
{special-NLPT-2}), which determine the\ general form of the NLPT. We notice,
in fact, that Eqs.(\ref{linear transf-0}) although formally analogous to the
$4-$velocity transformation laws generated by LPT, i.e., Eqs.(\ref{TU-1}),
are actually peculiar. This feature is reflected also in the transformation
matrices $\mathcal{A}_{\nu }^{\mu }$ and $\mathcal{B}_{\nu }^{\mu }.$ In
fact, although in view of the discussion given above both $\mathcal{B}_{\nu
}^{\mu }(r)$ and $\mathcal{A}_{\nu }^{\mu }\left( r\right) $ are simply
local functions of $r\equiv \left\{ r^{\mu }\right\} $, it is obvious that
the transformations Eqs.(\ref{NLPT-1}) have manifestly a non-local character.

\section{6 - A sample case: solution of the TT-problem for diagonal metric
tensors}

As pointed out above the theory of special NLPT must in principle hold also
when the space-times $(\mathbf{Q}^{4},g)$ and $(\mathbf{Q}^{\prime
4},g^{\prime })\equiv (\mathbf{M}^{\prime 4},\eta )$\textbf{\ }have
different signatures. In particular, if $(\mathbf{Q}^{4},g)$ coincides with
a flat space-time, then it might still have in principle an arbitrary
signature.\ To clarify this important point we present in this section a
sample application.\ For definiteness, let us consider here a curved
space-time $(\mathbf{Q}^{4},g)$ which is diagonal when expressed in terms of
Cartesian coordinate. The following two possible realizations are considered

\begin{itemize}
\item A) diag$\left( g_{\mu \nu }\right) \equiv $diag$\left(
S_{0}(r),-S_{1}(r),-S_{2}(r),-S_{3}(r)\right) $.

\item B) diag$\left( g_{\mu \nu }\right) \equiv $diag$\left(
-S_{0}(r),S_{1}(r),-S_{2}(r),-S_{3}(r)\right) $.
\end{itemize}

In both cases here the functions $S_{\mu }(r)$ are assumed to be \emph{%
prescribed} real functions which are strictly positive for all $r\equiv
r^{\mu }\in (\mathbf{Q}^{4},g)$. Since by construction the Riemannian
distance $ds$ is left invariant by arbitrary NLPTs, it follows that in the
two cases either the differential identity%
\begin{eqnarray}
ds^{2} &=&S_{0}\left( dr^{0}\right) ^{2}-S_{1}\left( dr^{1}\right)
^{2}-S_{2}\left( dr^{2}\right) ^{2}-S_{3}\left( dr^{3}\right) ^{2}  \notag \\
&=&\left( dr^{\prime 0}\right) ^{2}-\left( dr^{\prime 1}\right) ^{2}-\left(
dr^{\prime 2}\right) ^{2}-\left( dr^{\prime 3}\right) ^{2}
\label{eq-diagonal-1}
\end{eqnarray}%
or%
\begin{eqnarray}
ds^{2} &=&S_{0}\left( dr^{0}\right) ^{2}-S_{1}\left( dr^{1}\right)
^{2}-S_{2}\left( dr^{2}\right) ^{2}-S_{3}\left( dr^{3}\right) ^{2}  \notag \\
&=&-\left( dr^{\prime 0}\right) ^{2}+\left( dr^{\prime 1}\right) ^{2}-\left(
dr^{\prime 2}\right) ^{2}-\left( dr^{\prime 3}\right) ^{2}
\label{eq-diagonal-2}
\end{eqnarray}%
respectively must hold. Let us point out the solutions of the TT-problem,
i.e., Eq.(\ref{first equation}) or equivalent (\ref{last-equation}), in the
two cases.

\subsection{Solution of case A)}

In validity of Eq.(\ref{eq-diagonal-1}), if one adopts a special NLPT of the
form%
\begin{equation}
dr^{\mu }=\left( 1+A_{\left( \mu \right) }^{\left( \mu \right) }(r^{\prime
},r)\right) dr^{\prime \left( \mu \right) },  \label{Diagonal NLPT}
\end{equation}%
in terms of Eq.(\ref{last-equation}) this delivers for diagonal matrix
elements $A_{\left( \mu \right) }^{\mu }(r^{\prime },r)$ for all $\mu =0,3$
the equations%
\begin{equation}
1=S_{\mu }\left( r\right) \left( 1+A_{\left( \mu \right) }^{\left( \mu
\right) }(r^{\prime },r)\right) ^{2},
\end{equation}%
with the formal solutions%
\begin{equation}
A_{\left( \mu \right) }^{\mu }(r^{\prime },r)=\sqrt{\frac{1}{S_{\left( \mu
\right) }\left( r\right) }}-1.  \label{these}
\end{equation}%
Notice that here only the positive\ algebraic roots have been retained in
order to recover from Eq.(\ref{these}) the identity transformation when\
letting $S_{\mu }\left( r\right) =1$. \ From Eq.(\ref{special-NLPT-1}) one
obtains therefore the special NLPT%
\begin{equation}
r^{\mu }(s)=r^{\mu }(s_{o})+\int_{s_{o}}^{s}ds\frac{dr^{\prime \left( \mu
\right) }(s)}{ds}\sqrt{\frac{1}{S_{\left( \mu \right) }\left( r\right) }},
\label{FINAL-1}
\end{equation}%
where in the integrand $r$ is to be considered as an implicit function of $%
r^{\prime }$ and, as indicated above, $\frac{dr^{\prime \left( \mu \right)
}(s)}{ds}$ remains still arbitrary. Thus, explicit solution of Eq.(\ref%
{FINAL-1}) can be obtained by suitably prescribing $\frac{dr^{\prime \left(
\mu \right) }(s)}{ds}.$

\subsection{Solution of case B)}

Let us now consider the solution of the TT-problem when Eq.(\ref%
{eq-diagonal-2}) applies. For definiteness, let us look for a special NLPT
of the type:%
\begin{eqnarray}
dr^{0} &=&M_{\left( 1\right) }^{\left( 0\right) }(r^{\prime },r)dr^{\prime
\left( 1\right) }, \\
dr^{1} &=&M_{\left( 0\right) }^{\left( 1\right) }(r^{\prime },r)dr^{\prime
\left( 0\right) }, \\
dr^{2} &=&M_{\left( 2\right) }^{\left( 2\right) }(r^{\prime },r)dr^{\prime
\left( 2\right) }, \\
dr^{3} &=&M_{\left( 3\right) }^{\left( 3\right) }(r^{\prime },r)dr^{\prime
\left( 3\right) }.
\end{eqnarray}%
In terms of Eq.(\ref{last-equation}) this delivers for diagonal matrix
elements $M_{\left( \mu \right) }^{\left( \mu \right) }$ the equations%
\begin{eqnarray}
1 &=&S_{1}\left( r\right) M_{\left( 0\right) }^{\left( 1\right) }(r^{\prime
},r)^{2}, \\
1 &=&S_{0}\left( r\right) M_{\left( 1\right) }^{\left( 0\right) }(r^{\prime
},r)^{2}, \\
1 &=&S_{2}\left( r\right) M_{\left( 2\right) }^{\left( 2\right) }(r^{\prime
},r)^{2}, \\
1 &=&S_{3}\left( r\right) M_{\left( 3\right) }^{\left( 3\right) }(r^{\prime
},r)^{2},
\end{eqnarray}%
with the formal solutions%
\begin{eqnarray}
M_{\left( 0\right) }^{\left( 1\right) }(r^{\prime },r) &=&\sqrt{\frac{1}{%
S_{\left( 1\right) }\left( r\right) }}, \\
M_{\left( 1\right) }^{\left( 0\right) }(r^{\prime },r)^{2} &=&\sqrt{\frac{1}{%
S_{\left( 0\right) }\left( r\right) }}, \\
M_{\left( 2\right) }^{\left( 2\right) }(r^{\prime },r)^{2} &=&\sqrt{\frac{1}{%
S_{\left( 2\right) }\left( r\right) }}, \\
M_{\left( 3\right) }^{\left( 3\right) }(r^{\prime },r)^{2} &=&\sqrt{\frac{1}{%
S_{\left( 3\right) }\left( r\right) }}.
\end{eqnarray}%
Hence, the corresponding NLPT in integral from are found to be in this case:%
\begin{eqnarray}
r^{0}(s) &=&r^{0}(s_{o})+\int_{s_{o}}^{s}ds\frac{dr^{\prime \left( 1\right) }%
}{ds}\sqrt{\frac{1}{S_{\left( 0\right) }\left( r\right) }}, \\
r^{1}(s) &=&r^{1}(s_{o})+\int_{s_{o}}^{s}ds\frac{dr^{\prime \left( 0\right) }%
}{ds}\sqrt{\frac{1}{S_{\left( 1\right) }\left( r\right) }},
\end{eqnarray}%
\begin{eqnarray}
r^{2}(s) &=&r^{2}(s_{o})+\int_{s_{o}}^{s}ds\frac{dr^{\prime \left( 2\right) }%
}{ds}\sqrt{\frac{1}{S_{\left( 2\right) }\left( r\right) }}, \\
r^{2}(s) &=&r^{2}(s_{o})+\int_{s_{o}}^{s}ds\frac{dr^{\prime \left( 3\right) }%
}{ds}\sqrt{\frac{1}{S_{\left( 2\right) }\left( r\right) }},
\end{eqnarray}%
where, again, in the integrands $r$ is to be considered as an implicit
function of $r^{\prime }$ while $\frac{dr^{\prime \left( i\right) }}{ds}$
has to be suitably prescribed.

Cases A and B correspond respectively to curved space-times having the same
or different signatures with respect to the Minkowski flat space-time.%
\textbf{\ }Therefore, based on the discussion displayed above, it is
immediate to conclude that a NLPT which maps mutually the two space-times
indicated above\ must necessarily exist\ in all cases considered here.

\section{7 - \ Concluding remarks}

Physical insight on the class of transformations $\left\{ P_{s}\right\} $
denoted here as special non-local point transformations (NLPT)\ emerges from
the following two\ statements, represented respectively by: A) Proposition P$%
_{2}$\ of THM.1 and B) the explicit realization obtained by the $4-$velocity
transformation laws (\ref{linear transf-0}) which follows in turn from Eqs.(%
\ref{displacemtnts-NLPT}).

Let us briefly analyze the first one, i.e., in particular the fact that the
set $\left\{ P_{S}\right\} $\ is endowed with the structure of a group. For
this purpose, consider two arbitrary connected and time-oriented curved
space-times $(\mathbf{Q}_{(i)}^{4},g_{(i)})$ for $i=1,2$ and assume that the
corresponding admissible subsets of $(\mathbf{M}^{4},\eta ),$\ on which the
same space-times are mapped by means of special NLPT, have a non-empty
intersection with non-vanishing measure. The corresponding Jacobian are by
assumption of the type (\ref{Jacobinas-(i)}) so that their product must
necessarily belong to $\left\{ P_{S}\right\} $\ (Proposition P$_{2}$).%
\textbf{\ }The conclusion is of outmost importance from the physical
standpoint. Indeed, it implies that by means of two special NLPT it is
possible to mutually map in each other \emph{two, }in principle arbitrary,%
\emph{\ curved space-times}. Therefore, the same theory can be applied in
principle to the treatment of arbitrary curved space-times by means of the
establishment of corresponding functional connections in terms of products
of suitable special NLPT.

The validity of the second consideration is also of perspicuous evidence. In
fact, the geometry of the transformed space-time $(\mathbf{Q}^{4},g)$, which
is represented by its metric tensor $g_{\mu \nu }\left( r\right) $\ and the
corresponding Riemann curvature tensor $R_{\sigma \mu \nu }^{\rho }(r)$,\
specifically arises because of suitable non-uniform $4-$velocity
transformation laws prescribed here. These also give rise to a related
non-local point transformation (NLPT) occurring between the two space-times $%
(\mathbf{Q}^{4},g)$ and $(\mathbf{Q}^{\prime 4},g^{\prime })\equiv (\mathbf{M%
}^{4},\eta )$. In particular, in the case of the solution indicated above
\textbf{(see Section 6) }for the transformation matrix $\mathcal{B}_{\nu
}^{\mu }(r)$, it follows that the transformed $4-$velocity has the following
qualitative properties:

\begin{itemize}
\item Its time component, besides depending on the corresponding
time-component of the Minkowski space-time, in general may carry also finite
contributions which are linearly-dependent on all spatial components of the
Minkowskian $4-$velocity.

\item The spatial components of the same $4-$velocity depend linearly only
on the corresponding spatial components the Minkowskian $4-$velocity, and
hence remain unaffected by its time component, i.e., its energy content in
the Minkowski space-time.
\end{itemize}

In view of these considerations, we are now in position to draw the main
conclusions.

This investigation carried our in this paper concerns basic theoretical
issues and physical problems which have remained unsolved to date in the
literature and whose solution obtained here is unprecedented in the
literature and of critical importance in GR as well as relativistic theories
such as Classical Electrodynamics, Kinetic Theory, Fluid and Magnetofluid
Dynamics, Relativisic Quantum Mechanics. Indeed in this paper, a new
approach to the standard formulation of GR (SF-GR) has been investigated
based on\ the extension of\ the customary functional setting which lays at
the\ basis of the same SF-GR. As an application, the Einstein's Teleparallel
transformation problem (TT-problem) has been considered. This involves the
construction of a suitable invertible coordinate transformation which maps
in each other an arbitrary curved space-time $(\mathbf{Q}^{4},g)$ and the
flat Minkowski space-time $(\mathbf{Q}^{\prime 4},g^{\prime })\equiv (%
\mathbf{M}^{\prime 4},\eta )$, both assumed to be parametrized in terms of
Cartesian coordinates. As pointed out in THM.1, this requires necessarily
the introduction of a new type of NLPT. Unlike local point transformations
(LPT) traditionally adopted in SF-GR, these transformations have the
distinctive property that the transformed space-time $(\mathbf{Q}^{4},g)$
may exhibit a non-vanishing Riemann curvature tensor. In particular, the new
class of transformations is characterized by Jacobians with a characteristic
non-gradient form.

The present outcome implies the extension of the traditional concept of
point transformations and GR-reference frame adopted customarily in GR.\
Indeed, although the mathematical adequacy of SF-GR remains paramount in the
framework of the LPT-functional setting, we have shown that there exist
theoretical motivations based on actual physical problems\textbf{,} and in
particular the TT-problem arising in the context of Einstein's theory of
teleparallel gravity, which require a different functional setting. In fact,
the solution of the same TT-problem requires changing\ both the functional
setting as well as the notion of GR-reference frame usually adopted in GR.
More precisely:

\begin{enumerate}
\item Based on the prescription of the coordinates and corresponding $4-$%
velocity associated with each extended GR-frame, suitable phase-space
transformations among them, denoted as NLPT-phase-space transformations, are
introduced.

\item In particular, concerning the corresponding point transformations,
these are identified with the special NLPT $\left\{ P_{S}\right\} $\ (see
Eqs.(\ref{NLPT-1})).\ These transformations reduce locally to LPT if the
gradient conditions (\ref{RRPG-1})-(\ref{RRPG-2}) apply.

\item The coordinate systems mapped in each other by means of a special NLPT
belong, unlike in SF-GR, to two different space-times.\ In the case of the
TT-problem these have been identified respectively with a generic curved
space-time $(\mathbf{Q}^{4},g)$\textbf{\ }and the flat Minkowski\textbf{\ }%
space-time $(\mathbf{Q}^{\prime 4},g^{\prime })\equiv (\mathbf{M}^{\prime
4},\eta )$, both represented in terms of Cartesian coordinates.

\item The class of special NLPT includes also coordinate transformations
with map the Minkowski space-time onto a curved space-time characterized by
a different signatures.
\end{enumerate}

As shown here, the solution of\ the TT-problem rests purely on physical
principles. In this regard in the present paper the following remarks have
turned out to be crucial.

The first one is realized by Proposition \#1, namely the fact that two
different space-times, such as those occurring in the Einstein's
TT-problem,\ namely\textbf{\ }$(\mathbf{Q}^{4},g)$\ and $(\mathbf{Q}^{\prime
4},g^{\prime })\equiv (\mathbf{M}^{\prime 4},\eta )$,\ cannot be directly
mapped in each other just by means of a LPT.\textbf{\ }The second one, that
general $4-$velocity transformations of the form given by Eqs.(\ref{linear
transf-0}) manifestly can always be introduced in which the Jacobian of the
transformation is not of the gradient-form indicated by Eqs.(\ref{GR-1}) and
(\ref{GR-2}).\ The third fundamental remark concerns the existence of NLTP.
This is actually suggested by the Einstein equivalence principle itself, a
principle which also lies at the heart of his approach to the TT-problem.
Such a feature appears of critical importance. In fact, as shown here, it
directly leads to the identification of the precise form of the NLPT which
provides an explicit solution of the same TT-problem.

Finally, two characteristic aspects of the new (NLPT) transformations
proposed here must be stressed. The first one is their non-locality, which
appears both in their Lagrangian and Eulerian forms. This arises because of
their non-local dependence with respect to $4-$velocity. The second,\ and in
turn related, one\ is due to the form of their Jacobians. In fact, in
difference with the treatment of LPT, for NLPT the same ones are not
identified with gradient operators. Nevertheless, since the Jacobians still
are by assumption locally velocity-independent, tensor transformation laws
can actually once again be recovered. These follow from the corresponding
transformation equations which hold for the infinitesimal 4-position
displacements and the corresponding $4-$velocities.

\section{Acknowledgments}

Work developed within the research projects of the Czech Science Foundation
GA\v{C}R grant No. 14-07753P (C.C.)\ and Albert Einstein Center for
Gravitation and Astrophysics, Czech Science Foundation No. 14-37086G (M.T.).

\end{document}